\documentclass{aa}
\usepackage[varg]{txfonts}
\usepackage{natbib}
\usepackage{color}
\usepackage{rotating}
\usepackage{graphicx}
\bibpunct{(}{)}{;}{a}{}{,}

\begin{document}

\title{The ALMA Frontier Fields Survey}
\subtitle{IV: Lensing-corrected 1.1\,mm number counts in Abell 2744, MACSJ0416.1-2403 and MACSJ1149.5+2223}

\author{A. M. Mu\~noz Arancibia\inst{1} \and J. Gonz\'alez-L\'opez\inst{2,3} \and E. Ibar\inst{1} \and F. E. Bauer\inst{2,4,5} \and M. Carrasco\inst{6} \and N. Laporte\inst{7} \and T. Anguita\inst{8,4} \and M. Aravena\inst{3} \and F. Barrientos\inst{2} \and R. J. Bouwens\inst{9} \and R. Demarco\inst{10} \and L. Infante\inst{2,11} \and R. Kneissl\inst{12,13} \and N. Nagar\inst{10} \and N. Padilla\inst{2} \and C. Romero-Ca\~nizales\inst{14,3} \and P. Troncoso\inst{15,2} \and A. Zitrin\inst{16}}

\institute{Instituto de F\'isica y Astronom\'ia, Universidad de Valpara\'iso, Av. Gran Breta\~na 1111, Valpara\'iso, Chile\\e-mail: \texttt{alejandra.munozar@uv.cl}
\and Instituto de Astrof\'isica y Centro de Astroingenier\'ia, Facultad de F\'isica, Pontificia Universidad Cat\'olica de Chile, Casilla 306, Santiago 22, Chile
\and N\'ucleo de Astronom\'ia de la Facultad de Ingenier\'ia y Ciencias, Universidad Diego Portales, Av. Ej\'ercito 441, Santiago, Chile
\and Millennium Institute of Astrophysics, Chile
\and Space Science Institute, 4750 Walnut Street, Suite 205, Boulder, CO 80301, USA
\and Zentrum f\"{u}r Astronomie, Institut f\"{u}r Theoretische Astrophysik, Philosophenweg 12, 69120 Heidelberg, Germany
\and Department of Physics and Astronomy, University College London, Gower Street, London WC1E 6BT, UK
\and Departamento de Ciencias F\'isicas, Universidad Andres Bello, Av. Rep\'ublica 252, Santiago, Chile
\and Leiden Observatory, Leiden University, 2300 RA Leiden, The Netherlands
\and Department of Astronomy, Universidad de Concepci\'on, Casilla 160-C, Concepci\'on, Chile
\and Carnegie Institution for Science, Las Campanas Observatory, Casilla 601, Colina El Pino S/N, La Serena, Chile
\and Joint ALMA Observatory, Alonso de C\'ordova 3107, Vitacura, Santiago, Chile
\and European Southern Observatory, Alonso de C\'ordova 3107, Vitacura, Casilla 19001, Santiago, Chile
\and Chinese Academy of Sciences South America Center for Astronomy, National Astronomical Observatories, CAS, Beijing 100101, China
\and Universidad Aut\'onoma de Chile, Chile. Av. Pedro de Valdivia 425, Santiago, Chile
\and Physics Department, Ben-Gurion University of the Negev, P.O. Box 653, Be’er-Sheva 8410501, Israel
}

\date{}

\abstract{Characterizing the number counts of faint (i.e., sub-mJy and especially sub-$100\,\mu\textmd{Jy}$), dusty star-forming galaxies is currently a challenge even for deep, high-resolution observations in the FIR-to-mm regime. They are predicted to account for approximately half of the total extragalactic background light at those wavelengths. Searching for dusty star-forming galaxies behind massive galaxy clusters benefits from strong lensing, enhancing their measured emission while increasing spatial resolution. Derived number counts depend, however, on mass reconstruction models that properly constrain these clusters.}
{We aim to estimate the $1.1\,\textmd{mm}$ number counts along the line of sight of three galaxy clusters, Abell 2744, MACSJ0416.1-2403, and MACSJ1149.5+2223, which are part of the ALMA Frontier Fields Survey. We have performed detailed simulations to correct these counts for lensing effects, probing down to the sub-mJy flux density level.}
{We created a source catalog based on ALMA $1.1\,\textmd{mm}$ continuum detections. We used several publicly available lensing models for the galaxy clusters to derive the intrinsic flux densities of these sources. We performed Monte Carlo simulations of the number counts for a detailed treatment of the uncertainties in the magnifications and adopted source redshifts.}
{We estimate lensing-corrected number counts at $1.1\,\textmd{mm}$ using source detections down to $\textmd{S/N}=4.5$. In each cluster field, we find an overall agreement among the number counts derived for the different lens models, despite their systematic variations regarding source magnifications and effective areas. Combining all cluster fields, our number counts span $\sim2.5\,\textmd{dex}$ in demagnified flux density, from several mJy down to tens of $\mu\textmd{Jy}$. Both our differential and cumulative number counts are consistent with recent estimates from deep ALMA observations at a $3\sigma$ level. Below $\approx0.1\,\textmd{mJy}$, however, our cumulative counts are lower by $\approx1\,\textmd{dex}$, suggesting a flattening in the number counts.}
{We derive $1.1\,\textmd{mm}$ number counts around three well-studied galaxy clusters following a statistical approach. In our deepest ALMA mosaic, we estimate number counts for intrinsic flux densities $\approx4$ times fainter than the rms level. This highlights the potential of probing the sub-$10\,\mu\textmd{Jy}$ population in larger samples of galaxy cluster fields with deeper ALMA observations.}


\maketitle

\section{Introduction}

Observations at far-infrared (FIR) to millimeter (mm) wavelengths have revealed a population of dusty star-forming galaxies (DSFGs, see \citealt{Casey2014} and references therein). The detection of these sources benefits from the negative $k$-correction in their spectral energy distribution (SED), which keeps their measured flux density in the FIR-to-mm roughly constant up to redshift $z\approx6-10$ \citep{Blain2002}. Bright sources were first detected using single-dish telescopes (e.g., \citealt{Smail1997,Hughes1998}). After exhaustive identification efforts and spectroscopic campaigns, they were found to lie at high redshift with a peak at $z\sim2-2.5$ (e.g., \citealt{Chapman2005,Greve2005,Pope2006,Younger2007}).

The surface density of DSFGs detected at different wavelengths is quantified through galaxy number counts (e.g., \citealt{Blain1999c}). The bright end of the galaxy distribution has been extensively probed with single-dish telescopes (e.g., \citealt{Coppin2006,Weiss2009}). Recent interferometric follow-up observations of bright sources ($\gtrsim5\,\textmd{mJy}$ at $870\,\mu\textmd{m}$) have resolved some of them into multiple components \citep{Smolcic2012,Karim2013,Hodge2013}. Fainter DSFGs comprise the bulk of the star formation activity at high redshifts. It has been estimated that sources having $\simeq0.1-1\,\textmd{mJy}$ at $1.2\,\textmd{mm}$ account for $\gtrsim50\%$ of the total extragalactic background light (EBL) at mm wavelengths (e.g., \citealt{Ono2014,Carniani2015,Aravena2016,Fujimoto2016,Hatsukade2016,Hsu2016,Oteo2016}). Better constraints await a complete census of fainter galaxies at these wavelengths in order to fully understand the various contributions to the EBL. Importantly, measuring the source brightness at several FIR-to-mm bands helps to disentangle how the rest-frame FIR spectra vary among galaxy populations; this serves as a key constraint for models of galaxy formation and evolution (e.g., \citealt{Hayward2013,Cowley2015,MunozArancibia2015}).

Faint flux densities can be probed in two ways, namely 1) performing deeper, high-resolution observations (compared to current confusion-limited single-dish data), or 2) using strong gravitational lensing by massive galaxy clusters \citep{Hezaveh2011}. The high sensitivity of the Atacama Large Millimeter/submillimeter Array (ALMA) recently allowed the possibility to probe and characterize the faint end of the unlensed sub-mm population \citep{Ono2014,Carniani2015,Oteo2016,Hatsukade2016,Aravena2016,Dunlop2017}. On the other hand, the lensing power enhances the measured flux density of background sources and alleviates the effects of confusion \citep{Blain1999c}. Some of the very first single-dish detections were done in galaxy cluster fields \citep{Smail1997}. Number counts from single-dish detected sources behind galaxy clusters have successfully probed flux densities down to the sub-mJy level albeit with a statistical approach, since counterparts are not firmly known (e.g., \citealt{Knudsen2008,Zemcov2010,Johansson2011,Hsu2016}). Combining both approaches can maximize their benefits. For instance, \citet{Fujimoto2016} derived $1.2\,\textmd{mm}$  number counts down to a flux density of $\sim0.02\,\textmd{mJy}$ ($\gtrsim4\sigma$), using proprietary and archival deep ALMA data that included 66 blank fields and one lensed galaxy cluster field.

In this work, we derive $1.1\,\textmd{mm}$ number counts using dedicated ALMA observations \citep[hereafter Paper I]{Gonzalez-Lopez2017} and recent publicly available lensing models. We exploit ALMA's unique capabilities to search for sources behind three well-studied galaxy clusters, which are part of the Frontier Fields survey (FFs, \citealt{Lotz2017}). This is a legacy project combining the power of gravitational lensing of massive clusters with extremely deep multiband HST and Spitzer imaging of six strong-lensing clusters and adjacent parallel fields. With the help of several detailed mass models for each galaxy cluster, we can harness the magnification power of these clusters to recover the intrinsic (i.e., ``delensed'') emission from background galaxies. In turn this may allow us to probe fainter flux densities when compared to observations from blank field surveys. Combining several cluster fields also helps to reduce the impact of cosmic variance, that is, the field-to-field variation found in the volume density of sources due to large scale structure \citep{Trenti2008}.

This paper is organized as follows. $\S$\ref{sect_data} briefly describes the observational $1.1\,\textmd{mm}$ data and public lensing models used in this work. $\S$\ref{sect_method} details the methodology used to derive the number counts, including a careful treatment of the uncertainties in magnification for a given lens model, source position and adopted redshift. $\S$\ref{sect_results} presents our derived demagnified $1.1\,\textmd{mm}$ counts and places them in context compared to recent estimates from deep ALMA observations. $\S$\ref{sect_concl} summarizes our main findings. Throughout this paper, we adopt a flat $\Lambda$CDM cosmology with parameters $H_0=70\,\textmd{km}\,\textmd{s}^{-1}\textmd{Mpc}^{-1}$, $\Omega_m=0.3$ and $\Omega_{\Lambda}=0.7$, in order to match the cosmology for which the lens models were produced.

\section{Data}\label{sect_data}

\subsection{Observations with ALMA}\label{sect_obs}

\subsubsection{High-significance detections}\label{sect_snr5}

The sources used in this work are drawn from the individual ALMA $1.1\,\textmd{mm}$ detections in three of the galaxy clusters that comprise the FF survey, namely, Abell 2744 ($z=0.308$), MACSJ0416.1-2403 ($z=0.396$), and MACSJ1149.5+2223 ($z=0.543$), hereafter A2744, MACSJ0416, and MACSJ1149, respectively. They were observed as part of the ALMA Frontier Fields Survey (cycle 2 project \#2013.1.00999.S, PI: F. Bauer). Paper I introduces the $1.1\,\textmd{mm}$ mosaic images, data reduction and analy\-sis for these galaxy clusters. Each field covers an observed area of $\sim4.6\,\textmd{arcmin}^2$, and thus sum to a total image-plane area of $\sim14\,\textmd{arcmin}^2$. This corresponds to $\sim3$ times the area of the Hubble Ultra Deep Field (HUDF, \citealt{Dunlop2017}) and $\sim14$ times the initial ALMA Spectroscopic Survey in the HUDF (ASPECS, \citealt{Walter2016,Aravena2016}). With natural weighting, our continuum data reach rms depths of $\sim55-71\,\mu\textmd{Jy}\,\textmd{beam}^{-1}$ and have synthesized beam sizes between $\sim0\farcs5-1\farcs5$. A2744 was partially observed in a more extended configuration compared to the other cluster fields, leading to a longer mean projected baseline. As a result, A2744 achieves the highest resolution among these fields (see Paper I for details).

For each cluster field, we take into account the primary beam (PB) correction. The source extraction is done within the region where $\textmd{PB}>0.5$, that is, where the PB sensitivity is at least $50\%$ of the peak sensitivity. Sources are detected by searching for pixels with signal-to-noise ratio $(\textmd{S/N})\geq5$, which are then grouped as individual sources using the DBSCAN python algorithm \citep{Pedregosa2012}. In the following, we refer to the source S/N as the ratio of the peak intensity and the background rms. We note that depending on the spatial PB correction, sources having the same S/N may have different PB-corrected peak intensities. Unless noted, in the following we refer to source flux densities and peak intensities using PB-corrected values.

At $\textmd{S/N}\geq5$, we detect seven sources in A2744, four in MACSJ0416 and one in MACSJ1149. Since some sources appear to be resolved, we measure their integrated flux densities performing two-dimensional elliptical Gaussian fits in the \textit{uv}-plane using the UVMCMCFIT python algorithm \citep{Bussmann2016}. These fits also deliver the centroid position and size parameters for each source. Before applying lensing corrections, detected sources have peak intensities in the range $\sim0.33-1.43\,\textmd{mJy}\,\textmd{beam}^{-1}$, integrated flux densities in the range $\sim0.41-2.82\,\textmd{mJy}$, effective radii in the range $\lesssim0\farcs05-0\farcs37$ and axial ratios in the range $\sim0.17-0.66$. All of these sources have near-infrared (NIR) detected counterparts (based on deep HST F160W imaging). None of them are members of a FF cluster. We refer the reader to Paper I for more details regarding the source extraction procedure, the choice of the \textit{uv}-plane for estimating integrated source flux densities and sizes, and the search for NIR counterparts.

\begin{table*}
\begin{center}
\caption{Continuum detections at $\textmd{S/N}\geq4.5$.}
\begin{tabular}{ccccccc}
\hline \hline
ID & $\textmd{RA}_{\textmd{J2000}}$ & $\textmd{Dec}_{\textmd{J2000}}$ & S/N & $S_{1.1\,\textmd{mm,peak}}$ & $S_{1.1\,\textmd{mm,}uv\textmd{-fit}}$ & $z$\\
& [hh:mm:ss.ss] & [$\pm$dd:mm:ss.ss] & & $[\textmd{mJy}\,\textmd{beam}^{-1}]$ & $[\textmd{mJy}]$ &\\ \hline
A2744-ID01\tablefootmark{a} & 00:14:19.80 & -30:23:07.66 & 25.9 & $1.433\pm0.056$ & $1.570\pm0.073$ & 2.9\tablefootmark{c}\\
A2744-ID02\tablefootmark{a} & 00:14:18.25 & -30:24:47.47 & 14.4 & $1.292\pm0.091$ & $2.816\pm0.229$ & 2.482\tablefootmark{c}\\
A2744-ID03\tablefootmark{a} & 00:14:20.40 & -30:22:54.42 & 13.9 & $0.798\pm0.058$ & $1.589\pm0.125$ & $2.52_{-0.45}^{+0.23}$\tablefootmark{d}\\
A2744-ID04\tablefootmark{a} & 00:14:17.58 & -30:23:00.56 & 13.8 & $0.932\pm0.068$ & $1.009\pm0.074$ & $1.02_{-0.09}^{+0.32}$\tablefootmark{d}\\
A2744-ID05\tablefootmark{a} & 00:14:19.12 & -30:22:42.20 & 7.7 & $0.655\pm0.086$ & $1.113\pm0.135$ & $2.01_{-0.16}^{+0.69}$\tablefootmark{d}\\
A2744-ID06\tablefootmark{a} & 00:14:17.28 & -30:22:58.60 & 6.5 & $0.574\pm0.089$ & $1.283\pm0.241$ & $2.08_{-0.08}^{+0.13}$\tablefootmark{d}\\
A2744-ID07\tablefootmark{a} & 00:14:22.10 & -30:22:49.67 & 6.2 & $0.455\pm0.074$ & $0.539\pm0.082$ & $1.85_{-0.14}^{+0.16}$\tablefootmark{d}\\
A2744-ID08\tablefootmark{b} & 00:14:24.73 & -30:24:34.20 & 4.8 & $0.270\pm0.056$ & \dots & \dots\\
A2744-ID09\tablefootmark{b} & 00:14:21.23 & -30:23:28.70 & 4.7 & $0.256\pm0.055$ & \dots & \dots\\
A2744-ID10\tablefootmark{b} & 00:14:17.72 & -30:23:02.25 & 4.5 & $0.286\pm0.063$ & \dots & \dots\\
A2744-ID11\tablefootmark{b} & 00:14:22.63 & -30:23:30.45 & 4.5 & $0.253\pm0.056$ & \dots & \dots\\
MACSJ0416-ID01\tablefootmark{a} & 04:16:10.79 & -24:04:47.53 & 15.4 & $1.010\pm0.066$ & $1.319\pm0.103$ & 2.086\tablefootmark{c}\\
MACSJ0416-ID02\tablefootmark{a} & 04:16:06.96 & -24:03:59.96 & 6.8 & $0.406\pm0.062$ & $0.574\pm0.132$ & 1.953\tablefootmark{c}\\
MACSJ0416-ID03\tablefootmark{a} & 04:16:08.81 & -24:05:22.58 & 5.8 & $0.389\pm0.067$ & $0.411\pm0.072$ & $1.29_{-0.39}^{+0.11}$\tablefootmark{d}\\
MACSJ0416-ID04\tablefootmark{a} & 04:16:11.67 & -24:04:19.44 & 5.1 & $0.333\pm0.066$ & $0.478\pm0.166$ & $2.27_{-0.61}^{+0.17}$\tablefootmark{d}\\
MACSJ0416-ID05\tablefootmark{b} & 04:16:10.52 & -24:05:04.77 & 4.6 & $0.302\pm0.066$ & \dots & 1.849\tablefootmark{e}\\
MACSJ1149-ID01\tablefootmark{a} & 11:49:36.09 & +22:24:24.60 & 5.9 & $0.442\pm0.074$ & $0.579\pm0.134$ & 1.46\tablefootmark{c}\\
MACSJ1149-ID02\tablefootmark{b} & 11:49:40.32 & +22:24:42.00 & 4.6 & $0.524\pm0.113$ & \dots & \dots\\
MACSJ1149-ID03\tablefootmark{b} & 11:49:35.41 & +22:23:38.60 & 4.5 & $0.326\pm0.072$ & \dots & \dots\\
\hline
\end{tabular}
\tablefoot{Column 1: Source ID. Columns 2, 3: Centroid J2000 position of ID. Column 4: Signal-to-noise of the detection. Column 5: PB-corrected peak intensity and $1\sigma$ error. Column 6: PB-corrected integrated flux density and $1\sigma$ error from $uv$ fitting. Column 7: Source redshift.
\tablefoottext{a}{High-significance ($\textmd{S/N}\geq5$) detections. Already reported in Paper I.}
\tablefoottext{b}{Low-significance ($4.5\leq\textmd{S/N}<5$) detections. Instead of performing a $uv$ fitting, we estimate the integrated flux density using the peak intensity and assuming given source size parameters (see $\S$\ref{sect_snr45}). Since all but one of them lack clear counterparts (partly due to contamination) and spectroscopic redshifts, nor were they included in Paper II study, we assume a Gaussian redshift distribution with mean $z=2$ and $\sigma=0.5$ for these sources.}
\tablefoottext{c}{Spectroscopic redshift from GLASS, already noted in Paper II.}
\tablefoottext{d}{Photometric redshift found in Paper II. Best fit value and $1\sigma$ error from SED fitting are presented here only for reference, as we use the full probability distribution found for each photometric redshift.}
\tablefoottext{e}{Spectroscopic redshift from GLASS.}
}
\label{tab_snr45gals}
\end{center}
\end{table*}

\subsubsection{Going to fainter flux densities: $4.5\leq\textmd{S/N}<5$}\label{sect_snr45}

We push below the $\textmd{S/N}\geq5$ threshold of the 12 detections already reported in Paper I (and reintroduced in $\S$\ref{sect_snr5}) in order to extract more information from the maps contributing to the number counts. We decide to include all sources having $\textmd{S/N}\geq4.5$ in the natural-weighted mosaics, being extracted through the same procedure as high-significance detections. This adds four sources to A2744, one to MACSJ0416, and two to MACSJ1149. Although the fraction of spurious sources increases for all fields as we move to lower S/N values, we can statistically correct the counts for this effect.

Table \ref{tab_snr45gals} lists these low-significance detections, together with the high-significance detections from Paper I. Peak intensities of $4.5\leq\textmd{S/N}<5$ sources range from $\sim0.25$ to $\sim0.52\,\textmd{mJy}\,\textmd{beam}^{-1}$. Since a two-dimensional Gaussian fit in the $uv$-plane gives a highly uncertain measure of the integrated source flux density at low S/N, we use the peak intensity of the detections for estimating of the integrated flux densities as follows. For all our low-significance sources, we adopt as their observed effective radius and axial ratio the median values found for the high-significance sources, namely, $r_{\textmd{eff,obs}}=0\farcs23$ and $q_{\textmd{obs}}=0.58$ (see Paper I). Assuming this source size is consistent with \citet{Fujimoto2017} values. From source injection simulations (see $\S$\ref{sect_comp}), we find a typical ratio between the peak and integrated flux density for these size parameters of 0.85, 0.96, and 0.96 in A2744, MACSJ0416, and MACSJ1149, respectively. Scaling the peak intensities by these ratios, the integrated flux densities of the $4.5\leq\textmd{S/N}<5$ detections range from $\sim0.30$ to $\sim0.55\,\textmd{mJy}$. For estimating the centroid coordinates of each source, we take the $\textmd{S/N}\geq4.5$ pixel that established the detection, plus all surrounding pixels having $\textmd{S/N}\geq4$. We collect the coordinates of these pixels, obtain the median right ascension and declination among all of them and set these median values as estimates of the source centroid coordinates.

Including these detections, our final catalog is comprised by 19 sources. We highlight that none of the $4.5\leq\textmd{S/N}<5$ sources are part of the lists of lensed galaxies used by the lens modeling teams, therefore they do not influence to the lens models.

\subsection{Source redshifts}\label{sect_z}

In a galaxy cluster field, the observed magnification by gravitational lensing of a background source varies with both its relative position and redshift. Since we have accurate positions and deep HST imaging, we thus consider available spectroscopic and photometric redshift information.

\citet[hereafter Paper II]{Laporte2017} determine photometric redshifts for all our $\textmd{S/N}>5$ detections via SED fitting, finding a mean redshift of $z=1.99\pm0.27$. Five of these high-significance sources (A2744-ID01, A2744-ID02, MACSJ0416-ID01, MACSJ0416-ID02, and MACSJ1149-ID03) have spectroscopic redshifts from the GLASS survey \citep{Treu2015}, which are consistent with the photometric redshifts found. We refer the reader to Paper II for more details regarding the multiwavelength data used, photometry estimates and SED-fitting procedure.

We search for counterparts to our $4.5\leq\textmd{S/N}<5$ sources in several public catalogs reporting photometric and spectroscopic redshift estimates, including: photometric redshifts estimated by the CLASH team \citep{Postman2012,Molino2017} and the ASTRODEEP survey \citep{Castellano2016,DiCriscienzo2017}; catalogs of spectroscopic redshifts by \citet{Owers2011}, \citet{Ebeling2014}, \citet{Jauzac2016}, \citet{Kawamata2016}, \citet{Treu2016}, \citet{Mahler2018}, the GLASS survey \citep{Hoag2016}, and the CLASH survey using VIMOS \citep{Balestra2016} and MUSE \citep{Grillo2016,Caminha2017} at VLT; and redshift estimates for Herschel detections \citep{Rawle2016}. We find that only MACSJ0416-ID05 has a counterpart within $\approx0\farcs3$ with a secure spectroscopic redshift $z=1.849$. This was measured from NIR spectra as part of the GLASS survey, confirmed by fitting the continuum grism spectra to SED templates. This galaxy also has extensive multiwavelength broadband data from ASTRODEEP and CLASH.

For the remaining $4.5\leq\textmd{S/N}<5$ sources, all galaxies in the aforementioned catalogs having reliable redshift estimates are beyond $\approx1''$ from ALMA peak positions. In a few cases, these are contaminated by foreground sources. This makes identification of likely faint NIR emission particularly challenging, thus it is hard to gauge the veracity of these sources.

The choice of source redshifts is as follows. First, we use the spectroscopic redshifts for the five $\textmd{S/N}>5$ and one $4.5\leq\textmd{S/N}<5$ detections, respectively. These are presented in Table \ref{tab_snr45gals}. For the remaining $\textmd{S/N}>5$ sources, we use the photometric redshift probability distributions obtained in Paper II. In the aforementioned table, best fit values and $1\sigma$ errors from these distributions are presented for reference.

For sources lacking any redshift information (i.e., all but one $4.5\leq\textmd{S/N}<5$ sources), we assume a Gaussian redshift distribution centered at $z=2$ with standard deviation 0.5. This assumption is supported by the mean photometric redshift found in Paper II for the $\textmd{S/N}>5$ sources and by results from the literature found in blind mm detections reaching the sub-mJy level (e.g., \citealt{Aravena2016,Dunlop2017}). It is also consistent within $\approx1\sigma$ with the median redshift of dusty galaxies at $1.1\,\textmd{mm}$ predicted by \citet{Bethermin2015} using an empirical model, both including and not including strongly-lensed sources, for our chosen S/N threshold (assuming point sources).

\subsection{Lensing models}\label{sect_models}

\begin{table*}
\begin{center}
\caption{Lensing models considered in this work.}
\begin{tabular}{cc}
\hline \hline
Model & References\\ \hline
Caminha v4\tablefootmark{a} & \citet{Caminha2017}\\
CATS v4, v4.1 & \citet{Jullo2009,Richard2014,Jauzac2014,Jauzac2015b,Jauzac2016}\\
Diego v4, v4.1 & \citet{Diego2005,Diego2007,Diego2015}\\
GLAFIC v4\tablefootmark{b} & \citet{Oguri2010,Kawamata2016,Kawamata2018}\\
Keeton v4 & \citet{Keeton2010,Ammons2014,McCully2014}\\
Sharon v4 & \citet{Jullo2007,Johnson2014}\\
Williams v4 & \citet{Liesenborgs2006,Liesenborgs2007,Sebesta2016}\\
\hline
\end{tabular}
\tablefoot{All these models cover the region where our ALMA sources lie.
\tablefoottext{a}{Only available for MACSJ0416.}
\tablefoottext{b}{Only available for A2744 and MACSJ0416.}
}
\label{tab_models}
\end{center}
\end{table*}

A massive object (e.g., a galaxy cluster) deforms the space-time in its vicinity, acting as a gravitational lens (see \citealt{Kneib2011} for a review). In cluster fields, the light from background sources is deflected and magnified. Magnification estimates at each source position are essential for obtaining lensing-corrected flux densities and thus, the number counts. For this, we make use of gravitational lensing models produced by independent teams. Detailed explanations for the models (and their several versions) provided by each team can be found in the readme files publicly available in the FF website\footnote{http://archive.stsci.edu/prepds/frontier/lensmodels/}. In the following, different model versions from a given team are treated as separated models.

Each modeling team uses their own choice of assumptions and methods. Lensing mass inversion techniques include parametric, free-form (or ``non-parametric'') and hybrid (i.e., a mixture of both) models. Parametric models, as the name suggests, assume that the mass distribution can be represented by a superposition of analytical functions that depend on a limited number of free parameters. In most cases, these models are guided by the distribution of cluster members and their luminosities. Free-form models do not use this assumption, but find the solution directly from the multiple-image constraints (as a result, their resolution is often lower). Parametric models include Caminha \citep{Caminha2017}, CATS \citep{Jullo2009,Richard2014,Jauzac2014,Jauzac2015b,Jauzac2016}, GLAFIC \citep{Oguri2010,Kawamata2016,Kawamata2018}, Keeton \citep{Keeton2010,Ammons2014,McCully2014}, and Sharon \citep{Jullo2007,Johnson2014}. Williams \citep{Liesenborgs2006,Liesenborgs2007,Sebesta2016} is a free-form model, while Diego \citep{Diego2005,Diego2007,Diego2015} is hybrid. Brief descriptions of these and more models can be found in \citet{Coe2015} and \citet{Priewe2017}.

Table \ref{tab_models} lists the models considered in this work. These models are constrained by input archival observations (both from HST and ground based), redshifts and multiple image identifications. The reliability of these constraints has been collectively assigned by all teams, ranking each constraint as Gold, Silver or Bronze (see \citealt{Priewe2017}). Model versions v3 and newer are based on FF observations, with v4 and newer models using a considerably larger set of arcs and spectroscopic redshifts compared to previous versions. Model versions v4 and v4.1 vary in the set of constraints chosen by each team, with v4 models using only the most reliable constraints (i.e., images from the Gold sample\footnote{Note, however, that the choice of constraints for v4 models is not completely homogeneous across teams. For instance, Sharon included also few Silver and Bronze images in regions where the number of Gold images is small. Similarly, teams that released v4.1 versions added particular lower-ranked constraints: CATS added Silver images plus some very (photometrically) convincing candidates, while Diego added the full Silver and Bronze sets.}). For details regarding the selection of constraints and their reliability, we refer to the readme files publicly available for each lens model. We attempt to use the best data to date, so for all cluster fields we consider only v4 or newer models.

Lens models are comprised of maps of the normalized mass surface density $\kappa$ and shear $\gamma$ of the galaxy cluster, assuming a redshift $z=\infty$ background. The deflection field $\overrightarrow{\alpha}$ around the lensing object can be estimated from $\kappa$ as

\begin{equation}
\nabla\cdot\overrightarrow{\alpha}=2\kappa \label{eq_defl}
\end{equation}

\noindent \citep{Coe2008}. These maps are scaled to the source-plane $z$ of interest as

\begin{equation}
\kappa(z)=\kappa\frac{D_{LS}}{D_S},\quad\gamma(z)=\gamma\frac{D_{LS}}{D_S},\quad\overrightarrow{\alpha}(z)=\overrightarrow{\alpha}\frac{D_{LS}}{D_S},
\end{equation}

\noindent where the angular diameter distances $D_S$ and $D_{LS}$ are computed from source to observer and source to lens respectively. The magnification map at a given source-plane $z$ is obtained as (see \citealt{Coe2015})

\begin{equation}
\mu(z)=\frac{1}{|(1-\kappa(z))^2-\gamma(z)^2|}. \label{eq_mu}
\end{equation}

For each release, teams provide a lens model coined as ``best'', plus a range of individual reconstructions (hereafter the ``range'' maps) that sample the range of uncertainties, that is, there is one $\kappa$ and $\gamma$ map for each realization. The field of view and angular resolution adopted for presenting the maps, as well as the number of realizations provided, may vary across teams and model versions.

We use the full set of mass reconstructions for estimating uncertainties in both source magnifications and effective source-plane areas in a given lens model. These in turn are propagated to the number counts as explained in $\S$\ref{sect_method}. In order to use the models, ``range'' maps for $\kappa$ and $\gamma$ are reprojected to the size and resolution of the ALMA maps using a first order interpolation. Based on these, we obtain magnification maps for a given source redshift using Eq. \ref{eq_mu}, and deflect these maps (together with the PB-corrected rms maps) to the source plane using the deflection fields; if several pixels in the image plane are deflected to only one in the source plane, only the image-plane pixel with the highest magnification is kept and assigned to the source-plane pixel (following \citealt{Coe2015}). This is needed as effective areas are measured in the source plane. However, we adopt redshift probability distributions for most of the detections (see $\S$\ref{sect_z}), and therefore need to create source-plane maps for several redshift values. In order to make our Monte Carlo simulations faster at this step, we precompute source-plane maps for a fixed grid in redshift, using steps of $\Delta z=0.2$ in the range $z_{\textmd{min}}=0.4$ to $z_{\textmd{max}}=4$ for A2744 and MACSJ0416 ($z_{\textmd{min}}=0.6$ for MACSJ1149, given the higher cluster redshift). It is safe to consider only this redshift range since it contains all the adopted spectroscopic redshifts; also, all our photometric redshift distributions have zero values at $z\geq4$, and this limit is at $4\sigma$ from the mean redshift assumed for sources lacking redshift information.

When sampling the source redshift distributions across the Monte Carlo simulations, we also find (for each random $z$) the two closest values used in our set of precomputed source-plane maps, and use them for interpolating the effective source-plane areas at a given demagnified peak intensity. It is safe to use this approximation even for sources having spectroscopic redshifts, as we check that the predictions using their two closest redshift bins have no significant variation for most detections.

All v4 and v4.1 lens models cover the region where our detections lie. However, a fraction of the region where the ALMA maps have $\textmd{PB}>0.5$ is not fully covered by the GLAFIC v4 model ($\sim0.4\%$ for A2744) and the Williams v4 model ($\sim2\%$, $13\%$, and $5\%$ for A2744, MACSJ0416, and MACSJ1149, respectively). In these cases, we impose $\mu=1$ for the missing pixels in magnification maps, as their closest pixels have $\mu\approx1$. In total, we adopt for use eight, nine, and seven lens models for A2744, MACSJ0416, and MACSJ1149, respectively (see Table \ref{tab_models}). Since not all modelers provide deflection field maps for all realizations, we use Eq. \ref{eq_defl} to compute these maps from the $\kappa$ maps provided for the ``range'' models.

\section{Methodology}\label{sect_method}

We compute demagnified number counts at $1.1\,\textmd{mm}$. This requires recovering the demagnified (i.e., source-plane) integrated flux density $S_{\textmd{demag}}$ for each source, which is obtained as

\begin{equation}
S_{\textmd{demag}}=\frac{S_{\textmd{obs}}}{\mu}. \label{eq_fluxintr}
\end{equation}

\noindent Here, $S_{\textmd{obs}}$ corresponds to the measured (i.e., image-plane) integrated flux density and $\mu$ the source magnification (see $\S$\ref{sect_magnif}). We obtain the differential number counts at the $j$-th flux density bin as

\begin{equation}
\frac{dN}{d\log(S)}=\frac{1}{\Delta\log(S)}\sum\limits_{i}^{n}X_i, \label{eq_ndiff}
\end{equation}

\noindent where we sum the individual contribution $X_i$ to these counts by the sources that have demagnified flux densities within that bin. Similarly, we compute the cumulative number counts for the $k$-th flux limit as

\begin{equation}
N(>S_k)=\sum\limits_{i}^{m}X_i, \label{eq_ncumu}
\end{equation}

\noindent where we sum over the sources having $S_{\textmd{demag},i}\geq S_k$. In these two expressions, we estimate the contribution by the $i$-th source as

\begin{equation}
X_i=\frac{1-p_{\textmd{false},i}}{C_i\,A_{\textmd{eff},i}}. \label{eq_xis}
\end{equation}

\noindent Here, $C_i$ is a completeness correction (see $\S$\ref{sect_comp}) and $p_{\textmd{false},i}$ the fraction of spurious sources (see $\S$\ref{sect_fsp}). $A_{\textmd{eff},i}$ corresponds to the effective area where that source can be detected (see $\S$\ref{sect_areaeff}), depending on the source redshift and lens model that is adopted.

A detailed treatment for all these quantities is described in this section. Throughout our number count analysis, we consider ALMA detections down to $\textmd{S/N}=4.5$. This S/N threshold is chosen as an appropriate balance between the correction factors that are related to the source detection, even when the fraction of spurious sources is not exactly the same among cluster fields (see $\S$\ref{sect_fsp}).

For the sake of simplicity, we assume that none of the ALMA continuum detections are multiply imaged over the S/N threshold. We verified this for all lens models using their ``best'' maps (see $\S$\ref{sect_models}). For each ALMA detection, we create a set of image-plane mosaic pixels, comprised by its peak plus all the $\textmd{S/N}\geq4$ pixels surrounding it (hereafter Set 1). For a given lens model and redshift, we use the deflection fields for finding the spatial coordinates of Set 1 pixels in the source plane. We later search for all the image-plane pixels that are deflected from these source-plane coordinates. This new set includes Set 1 pixels (by construction) but may include new mosaic positions if the source-plane pixels are multiply imaged. If any of these new pixels belongs to any of the remaining ALMA detections, that is, matches another peak pixel or a $\textmd{S/N}\geq4$ pixel surrounding it, a detection is said to be multiply imaged (above the S/N threshold) in our mosaics.

Adopting the same redshift bins as when precomputing magnification maps, we find that none of our $\textmd{S/N}\geq4.5$ detections are a multiple image of another one in the catalog. Moreover, we check that none of the newly found image-plane positions have $\textmd{S/N}\geq4$. Therefore, if any of our detections both lies at one of the redshifts considered and is multiply imaged, then the predicted images could not be detected, unless a $\textmd{S/N}<4$ criterion is used. We further assume that we have recovered the total $1.1\,\textmd{mm}$ flux densities, within their respective errors.

\subsection{Completeness}\label{sect_comp}

\begin{figure}
\centering
\resizebox{0.9\hsize}{!}{\includegraphics{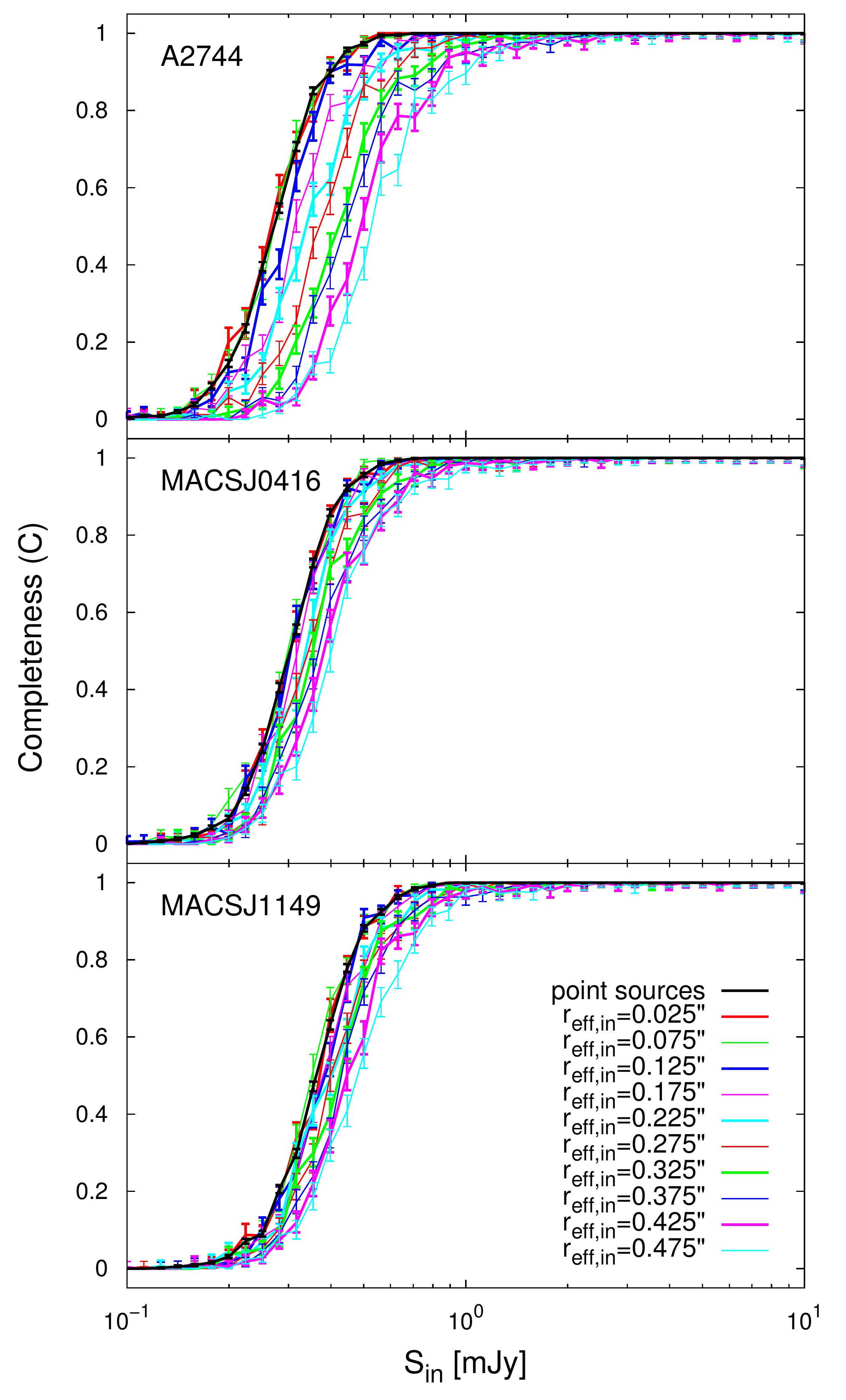}}
\caption{Completeness correction $C$ as a function of image-plane integrated flux density and separated in bins of image-plane scale radius. Error bars indicate binomial confidence intervals.}
\label{fig_comp_fluxout_reffin}
\end{figure}

\begin{figure}
\centering
\resizebox{0.9\hsize}{!}{\includegraphics{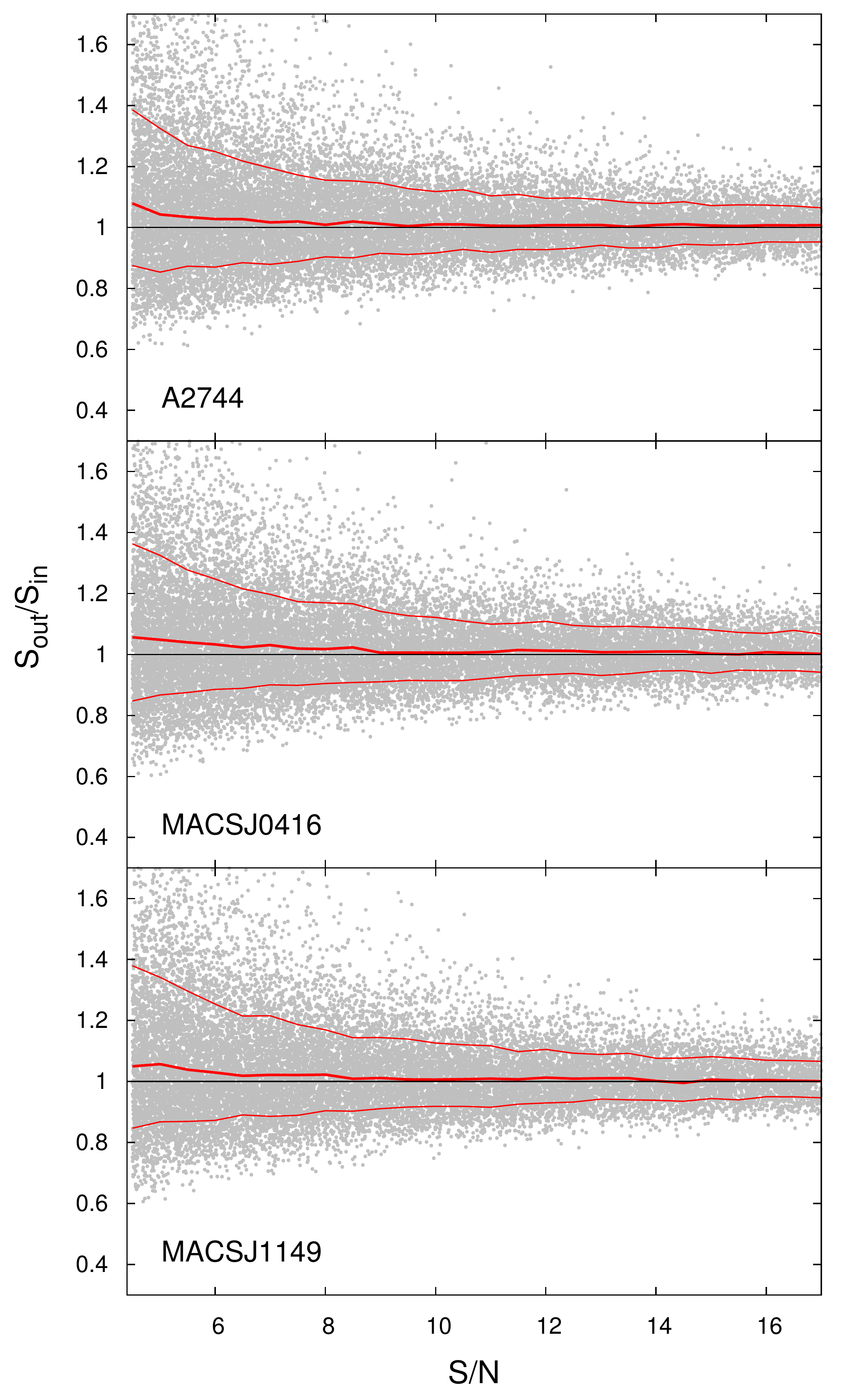}}
\caption{Deboosting correction as a function of S/N. We display the ratio between the extracted and injected flux densities for our simulated sources as gray dots. Thick red lines correspond to median values while thin red lines indicate the 16th and 84th percentiles.}
\label{fig_debcorr}
\end{figure}

In presence of noisy data, number counts need to be corrected for the proportion of sources that were not detected, because their noise level shifted their peak S/N below our chosen threshold. We compute the completeness as a function of image-plane integrated flux density $S_{\textmd{obs}}$ as follows. We draw $10^5$ artificial image-plane sources from a uniform distribution in $\log(S_{\textmd{obs}})$ in the range $0.01-10\,\textmd{mJy}$, a uniform distribution in scale radius $r_{\textmd{eff,obs}}$ in the range $0\farcs001-0\farcs5$ (sources with scale radii smaller than the pixel size are considered point sources) and a uniform distribution in axial ratio $q_{\textmd{obs}}$ in the range $0-1$. The scale radius interval is chosen based on the image-plane scale radii and their $1\sigma$ errors found for our high-significance detections. One at a time, we inject these sources randomly in the PB-corrected mosaic. We later extract them and check if they meet our $\textmd{S/N}\geq4.5$ criterion. We restrict this source injection only to the $\textmd{PB}>0.5$ region. We obtain the completeness $C$ for each (injected) flux bin as the fraction of sources that have an (extracted) $\textmd{S/N}\geq4.5$ and are thus detected. We later calculate the completeness curves assuming extended sources in steps of $\Delta r_{\textmd{eff,obs}}=0\farcs05$. The completeness corrections for all cluster fields are shown in Fig. \ref{fig_comp_fluxout_reffin}. For point sources, a value of $50\%$ is reached at image-plane flux densities of 0.27, 0.30, and $0.36\,\textmd{mJy}$ for A2744, MACSJ0416, and MACSJ1149, respectively. However, the completeness drops to $24\%$, $35\%$, and $42\%$ at the same flux densities for image-plane source sizes in the range $0\farcs20-0\farcs25$ (i.e., for the image-plane size assumed for our low-significance detections).

Since our source catalog is S/N limited, we note that measured source intensities may be systematically enhanced by noise fluctuations, such that they are boosted over the S/N threshold and thus bias the number counts. Correcting for this effect is known as flux deboosting (e.g., \citealt{Hogg1998,Weiss2009}). Taking the source injection simulations used to estimate the completeness corrections, we select the simulated sources extracted down to $\textmd{S/N}=4.5$ and compute the ratio between their extracted and injected flux densities. Figure \ref{fig_debcorr} shows these ratios, together with the median values found as a function of S/N. At $\textmd{S/N}=4.5$, we find that the noise boosts the flux densities by $8\%$, $6\%$, and $5\%$ for A2744, MACSJ0416, and MACSJ1149, respectively. We use the median ratios found at each S/N for correcting both the observed peak intensities and integrated flux densities for our detections.

If the underlying distribution of source flux densities is steep, number counts derived in the image plane can be overestimated even more in the faint end due to noise fluctuations. This is known as the Eddington bias \citep{Eddington1913}. Correcting the intrinsic number counts for this effect is not trivial, since it requires several assumptions to be made regarding the source properties and folding these through the various lens models. We choose to make no assumptions regarding the true underlying distribution of flux densities, supported by the low number density of ALMA sources in the FFs. However, we can obtain a rough estimate of the scope of any Eddington bias using a single ``trial'' lens model and assuming specific source flux density and redshift distributions. We choose to test this effect creating sets of $10^4$ simulated sources drawn from the redshift and flux distribution at 1.1mm predicted by the SIDES galaxy formation model \citep{Bethermin2017}, assuming random source coordinates, and lensing them using the ``best'' CATS v4 model. We then inject and extract these sources in our ALMA mosaics down to $\textmd{S/N}=4.5$, estimate the demagnified flux densities for these extracted sources and compute the ratio between output and input demagnified flux density as a function of S/N. At $\textmd{S/N}=4.5$, we estimate flux enhancements by $15\%$, $11\%$, and $13\%$ for A2744, MACSJ0416, and MACSJ1149, respectively. We find that within the uncertainties, these ratios are consistent with the deboosting corrections obtained in Fig. \ref{fig_debcorr}, which were computed assuming a flat distribution in $\log(S_{\textmd{obs}})$. We note that the counts predicted by the SIDES simulation agree with our demagnified counts at a $1\sigma$ level (see Fig. \ref{fig_countsdiffcumu_comp} and $\S$\ref{sect_counts_comp}). However, the SIDES simulation predicts steeper counts at flux densities $0.01-0.1\,\textmd{mJy}$ compared to our median estimates. Therefore, we consider that it is safe to skip any additional Eddington bias correction in this work, including only the deboosting correction shown in Fig. \ref{fig_debcorr}.

\subsection{Fraction of spurious sources}\label{sect_fsp}

\begin{figure}
\centering
\resizebox{0.9\hsize}{!}{\includegraphics{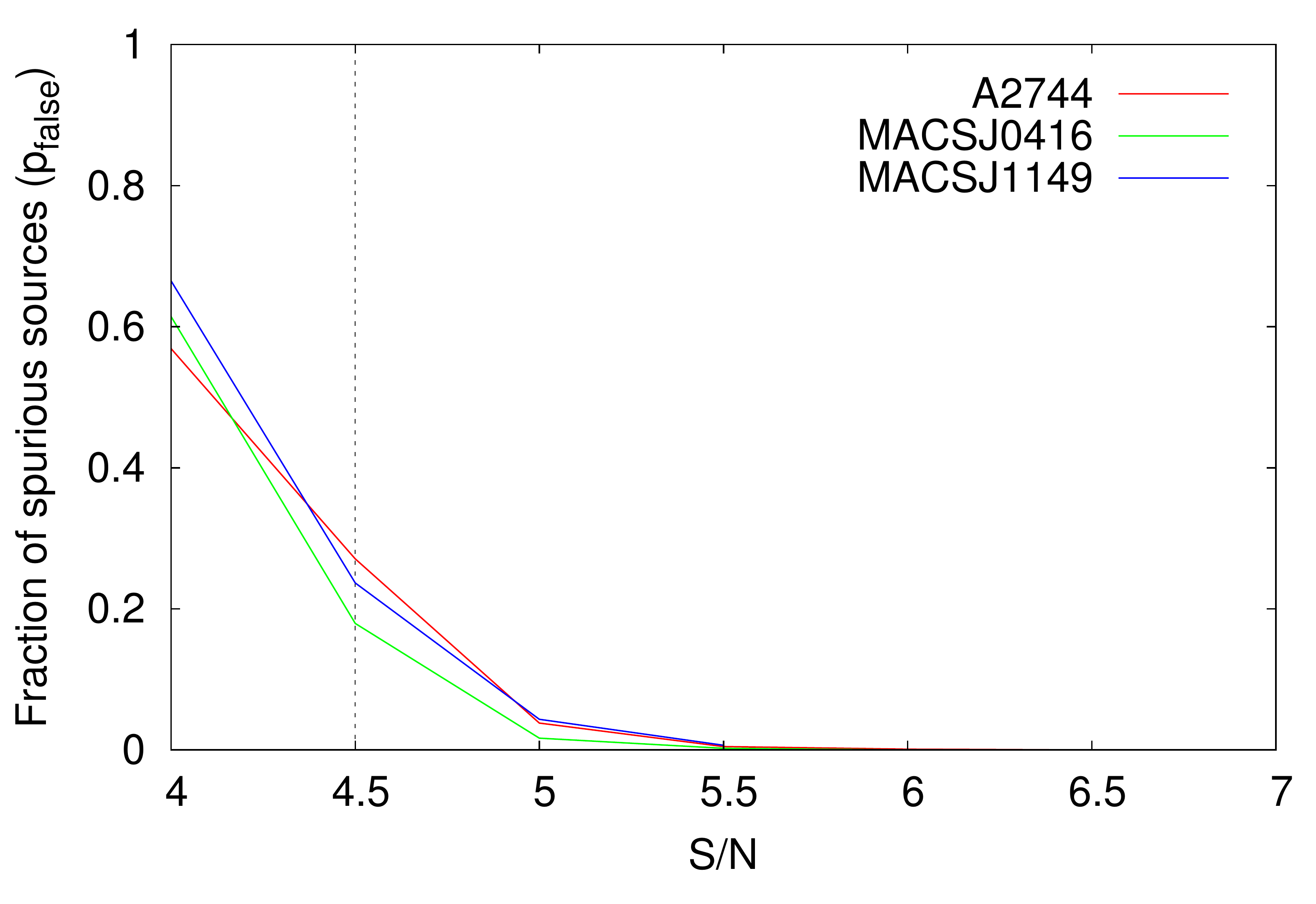}}
\caption{Fraction of spurious sources at a given S/N. We display curves for A2744, MACSJ0416, and MACSJ1149 in red, green, and blue, respectively. A vertical dotted line indicates our S/N threshold of 4.5.}
\label{fig_fsp}
\end{figure}

We compute the fraction of spurious sources (i.e., generated by noise) as a function of S/N as follows. For each galaxy cluster field, we generate 300 simulated non-PB-corrected maps, having the same size and resolution as the true ALMA mosaics. Each fake map is comprised by pure Gaussian noise with mean zero and variance one (in S/N units), convolved with the ALMA synthesized beam and later renormalized by the standard deviation of the noise distribution (for preserving the initial variance). We extract sources from each simulated map just as done with the true maps (see Paper I). Since the effective number of independent beams is twice the value expected from Gaussian statistics (see \citealt{Condon1997,Condon1998}), we double the number of sources detected in each noise map; doubling this number gives good agreement with the amount of sources found in the negative ALMA mosaics. We obtain the fraction of spurious sources at a given S/N, $p_{\textmd{false}}$, defined as the average ratio between the number of sources detected over that peak S/N in the true mosaic and in the simulated noise maps.

Figure \ref{fig_fsp} shows the fraction of spurious sources per $\textmd{S/N}$ limit for the three clusters. At $\textmd{S/N}\geq4.5$ $p_{\textmd{false}}$ is $\approx20\%-30\%$ among the cluster fields. Based on the source extraction on the 300 simulated noise maps, the average number of spurious sources at $\textmd{S/N}\geq4.5$ is $2.98\pm2.37$ (A2744), $0.90\pm1.30$ (MACSJ0416), and $0.81\pm1.32$ (MACSJ1149). This is consistent within $1\sigma$ with both the amount of spurious sources from the negative mosaics (five, one, and one, respectively) and the number of sources beyond $1''$ of an optical counterpart (four, zero, and two, respectively).

\subsection{Source magnifications}\label{sect_magnif}

\begin{figure*}
\centering
\resizebox{0.85\hsize}{!}
{\includegraphics{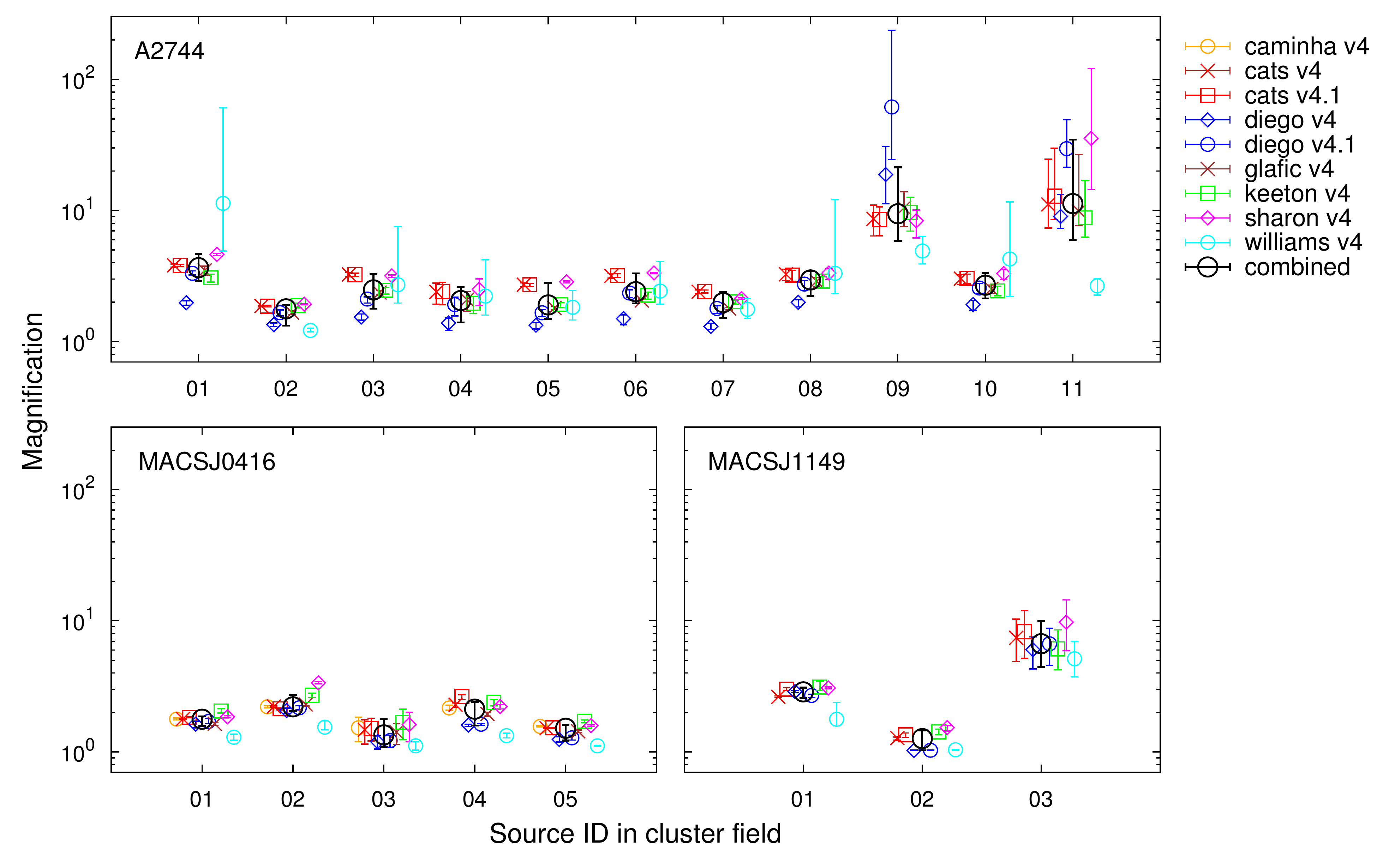}}
\caption{Median magnification per source for the lens models listed in Table \ref{tab_models} (colored symbols), and also combining all models for each cluster field (large black circles). Error bars indicate the 16th and 84th percentiles (see $\S$\ref{sect_magnif}). Values for each model have been offset around the source ID for clarity.}
\label{fig_magnif_gal}
\end{figure*}

\begin{figure*}
\centering
\resizebox{0.85\hsize}{!}
{\includegraphics{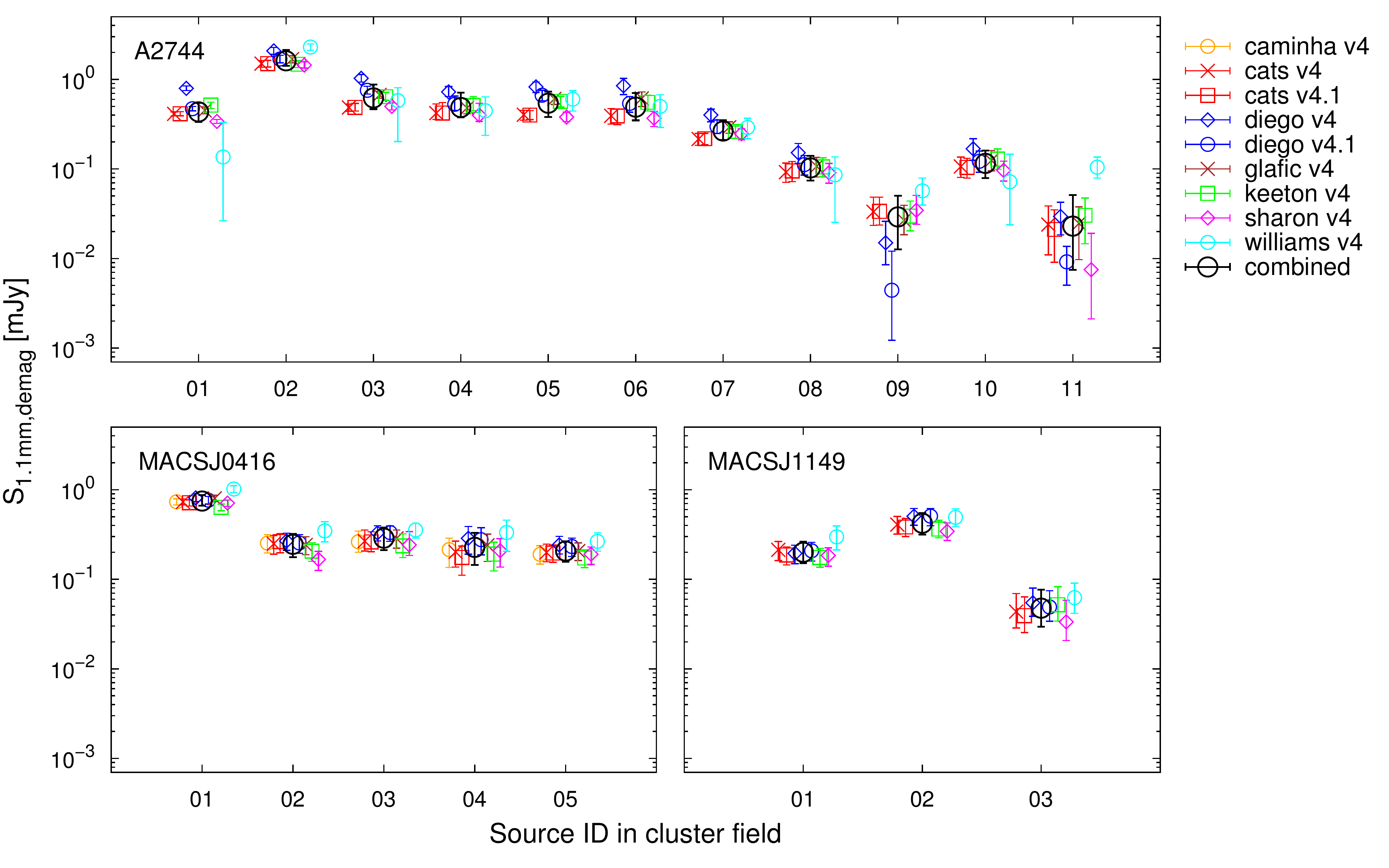}}
\caption{Median demagnified integrated flux density per source for the lens models listed in Table \ref{tab_models} (colored symbols), and also combining all models for each cluster field (large black circles). Error bars indicate the 16th and 84th percentiles. Values for each model have been offset around the source ID for clarity.}
\label{fig_flux_intr_gal}
\end{figure*}

Predicting how much is the source brightness amplified by the gravitational lensing effect is necessary for estimating the intrinsic emission from background sources. Lens models applying different techniques predict different values for that magnification.

The centroid pixel of each ALMA detection (see $\S$\ref{sect_obs}), together with the ``range'' maps (see $\S$\ref{sect_models}), are used to calculate the magnification for each source. Indeed, we obtain the magnification distribution for a given source and lens model using the $\mu$ values found for the source centroid pixel in all the ``range'' maps. This choice implies neglecting the effects of differential magnification, and is done in order to simplify the calculations. This is safe as most detections lie far from critical lines (i.e., where magnification formally diverges), and thus magnifications do not have a strong variation across the image-plane extension of these sources. A few detections are found close to critical lines (A2744-ID09 and A2744-ID11), being as close as $\approx1$ synthesized beam away from them in a limited number of lens models and assumed redshifts. Unfortunately, these sources lack redshift information, making it difficult to constrain their source magnification (see Fig. \ref{fig_magnif_gal}). Notably, the predictive power of lens models is lower close to critical lines (see below), and thus these sources have large uncertainties in their magnifications.

Since we are adopting a non-unique redshift, we use a Monte Carlo approach with 1000 realizations. Each time, we draw a random $z$ value from the source redshift probability distribution (see $\S$\ref{sect_z}), choose randomly one of the ``range'' sets of $\kappa$ and $\gamma$ maps, and obtain the corresponding $\mu$ value using Eq. \ref{eq_mu}. If a sample $z$ is lower than the cluster redshift (e.g., the photometric redshift distribution has a non-zero probability which extends below the cluster redshift), we assume $\mu=1$ for the source (i.e., the source is not affected by lensing at that redshift), use its observed flux density and compute the corresponding effective area in the image plane (i.e., assuming all map pixels have $\mu=1$). This happens only to sources A2744-ID03 and A2744-ID04 and at a very low rate ($\sim3\%$ and $<1\%$ of the realizations, respectively), thus the inclusion of photometric redshift tails below the cluster redshift has a negligible impact in our results.

The magnification distribution sampled for each source is then a combination of distributions obtained at the source position for several redshifts. From this sampling, we can compute a median magnification for each source and estimate uncertainties using the 16th and 84th percentiles (following \citealt{Coe2015}). This is shown in Fig. \ref{fig_magnif_gal} for the models listed in Table \ref{tab_models}, and also combining all models for each cluster field. Median (combined) magnification values for our sample range from 1.3 to 11.3.

In a given lens model, we find that sources having higher median magnifications have also larger dispersions. Some sources having median $\mu\gtrsim10$ reach dispersions $\gtrsim0.5\,\textmd{dex}$, such as sources A2744-ID09 in the Diego v4.1 model and A2744-ID11 in the Sharon v4 model. Magnification distributions are broad and asymmetrical for sources A2744-ID01, A2744-ID03, A2744-ID04, A2744-ID08, and A2744-ID10 in the Williams v4 model, although most of them have median $\mu<10$. Sources in MACSJ0416 have very similar magnifications in all models, showing small individual dispersions.

Previous works have used the lens models publicly available in the FFs for quantifying systematic uncertainties in predicted magnifications, applying the lens models both to observations (e.g., \citealt{Bouwens2017}, \citealt{Lotz2017}, \citealt{Priewe2017}) and simulations (e.g., \citealt{Johnson2016}, \citealt{Acebron2017}, \citealt{Meneghetti2017}). Our trend of increasing dispersion with source magnification (see Fig. \ref{fig_magnif_gal}) is in line with results by \citet{Zitrin2015}, \citet{Meneghetti2017} and \citet{Bouwens2017}. \citet{Zitrin2015} presented a comprehensive lensing analysis of the complete CLASH cluster sample, examining several lens models produced by their team. They found that the systematic differences (relative to one of the models) increase rapidly with the magnification value. \citet{Meneghetti2017} made a detailed comparison of the mass reconstruction techniques applied by different teams using two simulated galaxy clusters, which resemble the depth and resolution of the FFs. They found that the largest uncertainties in lens models are close to cluster critical lines, with the predictive power of the lens models worsening at $\mu>10$. For instance, they estimated that the accuracy in the magnifications predicted by some models degrades from $\sim10\%$ at $\mu=3$ to $\sim30\%$ at $\mu=10$. \citet{Bouwens2017} found similar results using a sample of 160 lensed, NIR-detected sources at $z\sim6$ in the first four FFs. They constrained the faint end of the $z\sim6$ ultraviolet luminosity function (UV LF), finding systematic variations in the LF of several orders of magnitude at $M_{\textmd{UV,AB}}=-12\,\textmd{mag}$ and fainter. They attributed this to the large systematic uncertainties inherent at high magnifications, with models having a poor predictive power specially at $\mu>30$. 

Furthermore, \citet{Lotz2017} computed method-to-method standard deviations for the subset of models in A2744 and MACSJ0416 that kept using the same methodology across versions (i.e., for both pre- and post-FF data). They found no significant reduction in the magnification variations across methodologies, reporting median systematic uncertainties in magnification of $<26\%$ and $15\%$, for v3 models in A2744 and MACSJ0416, respectively. However, \citet{Priewe2017} found a systematic uncertainty of $70\%$ at $\mu\sim40$, using the dispersion between v3 or newer lens models in those cluster fields for a $z=9$ source plane. They argued that the discrepancies in the magnification predictions among models, which often exceed the statistical uncertainties reported by individual reconstructions, were driven by lensing degeneracies, that is, different mass distributions may reproduce the same observational constraints. Moreover, they found the Williams v3 model gives the largest magnification uncertainties at most sky locations in A2744. The broad magnification distributions that we find for some sources in A2744 in the Williams v4 model (see Fig. \ref{fig_magnif_gal}) are in line with these findings.

\begin{figure}
\centering
\resizebox{0.85\hsize}{!}
{\includegraphics{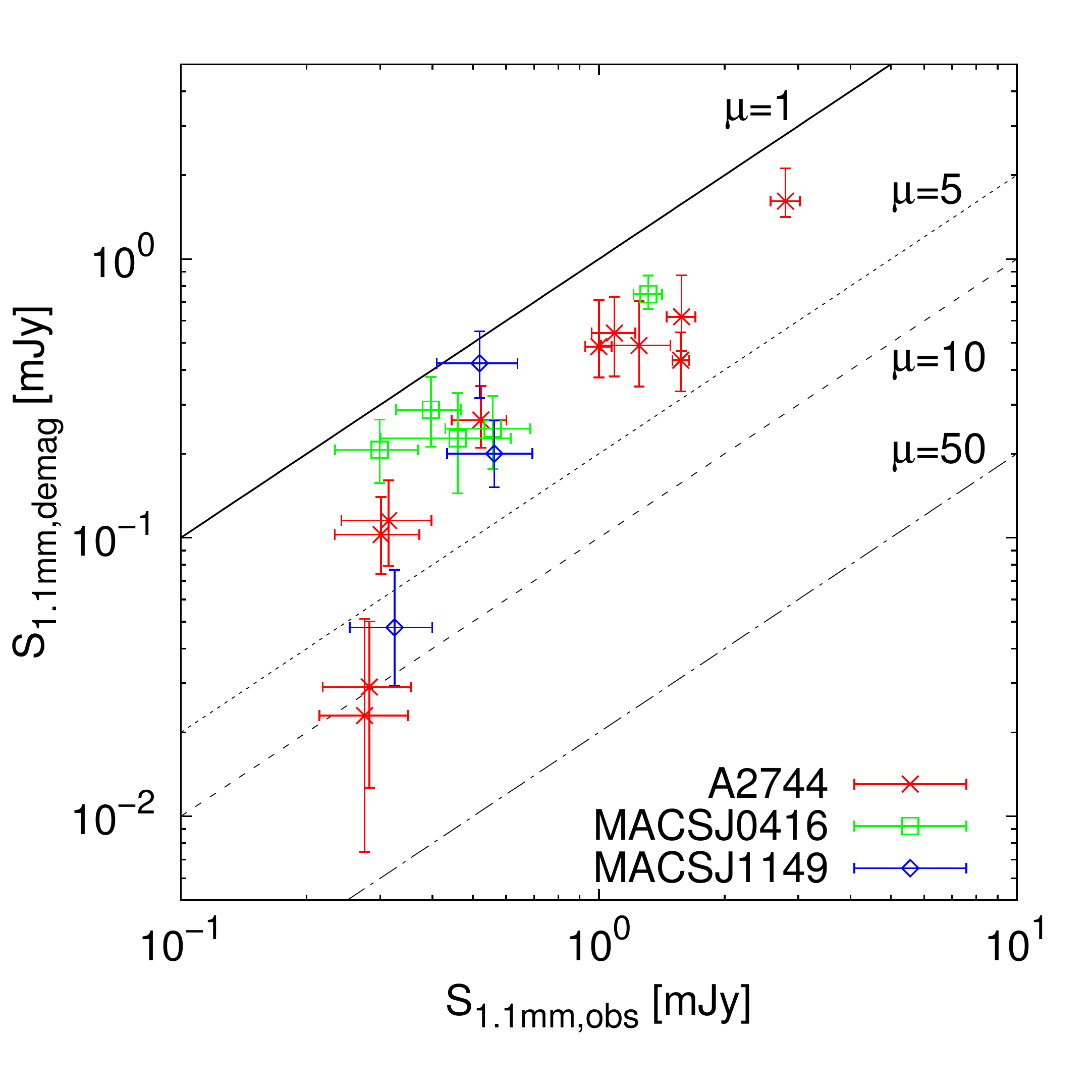}}
\caption{Median demagnified integrated flux density as a function of observed integrated flux density for A2744 (red crosses), MACSJ0416 (green squares), and MACSJ1149 (blue diamonds). Median values are obtained combining all models for each cluster field. Error bars in demagnified fluxes correspond to the 16th and 84th percentiles while for observed fluxes are $1\sigma$ statistical uncertainties. As a reference, black lines indicate magnification values of one (solid), five (dotted), ten (dashed) and 50 (dot-dashed).}
\label{fig_flux_intr_flux_obs}
\end{figure}

\subsection{Lensing-corrected source flux densities}\label{sect_fluxintr}

Once the magnification distribution for each source is obtained, the demagnified integrated flux density is recovered using Eq. \ref{eq_fluxintr} for the different lens models. We do this by adopting a Gaussian distribution for $S_{\textmd{obs}}$ with standard deviation given by its reported statistical error, and the distribution described in $\S$\ref{sect_magnif} for the magnification. Using both, we resample 1000 times the ratio given in Eq. \ref{eq_fluxintr} to obtain a distribution for $S_{\textmd{demag}}$.

Figure \ref{fig_flux_intr_gal} shows the median demagnified integrated flux density for each source, computed from both the distributions obtained for each model and joining all of them for each cluster field. Median (combined) lensing-corrected flux densities range from $\sim0.02$ to $1.62\,\textmd{mJy}$, with both the faintest and brightest sources in the sample being found around A2744. Naturally, sources having broad magnification distributions have also large uncertainties in their median $S_{\textmd{demag}}$ values. Within the uncertainties, combined demagnified flux densities cover around 2.5 orders of magnitude.

At $S_{\textmd{obs}}\gtrsim0.4\,\textmd{mJy}$, we find a trend of brighter observed sources being also brighter intrinsically, while sources having lower observed flux densities tend to span $\approx1.5\,\textmd{dex}$ in demagnified flux. This is shown in Fig. \ref{fig_flux_intr_flux_obs}. We also find that sources with the highest magnifications ($\mu\gtrsim5$) are among the faintest ones both in observed and lensing-corrected flux ($S_{\textmd{obs}}\lesssim0.4\,\textmd{mJy}$ and $S_{\textmd{demag}}\lesssim0.06\,\textmd{mJy}$, respectively).

\begin{figure}
\centering
\resizebox{0.9\hsize}{!}
{\includegraphics{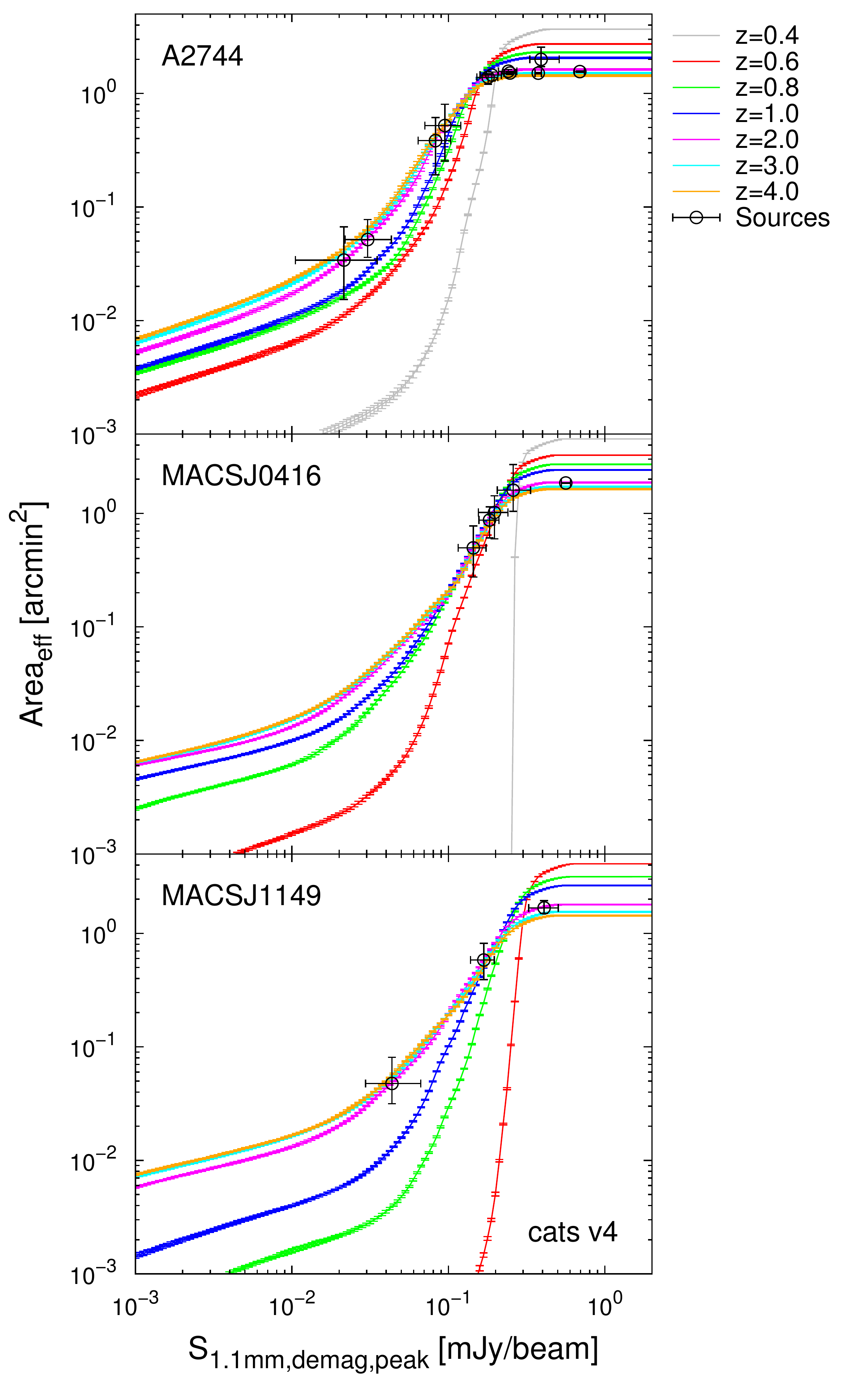}}
\caption{Median effective area as a function of demagnified peak intensity at several redshifts as indicated in the key (colored lines) for the CATS v4 lens model. Values for our $\textmd{S/N}\geq4.5$ sources (black symbols) are shown for this model (corresponding to the red crosses in Fig. \ref{fig_area_gal}). Error bars indicate the 16th and 84th percentiles. For each curve, these are obtained using the ``range'' maps at the corresponding redshift, while for symbols they are computed as described in $\S$\ref{sect_areaeff}. At $z\geq2$, areas do not differ significantly with redshift for this lens model.}
\label{fig_area_flux_z}
\end{figure}

\subsection{Source effective areas}\label{sect_areaeff}

\begin{figure*}
\centering
\resizebox{0.85\hsize}{!}
{\includegraphics{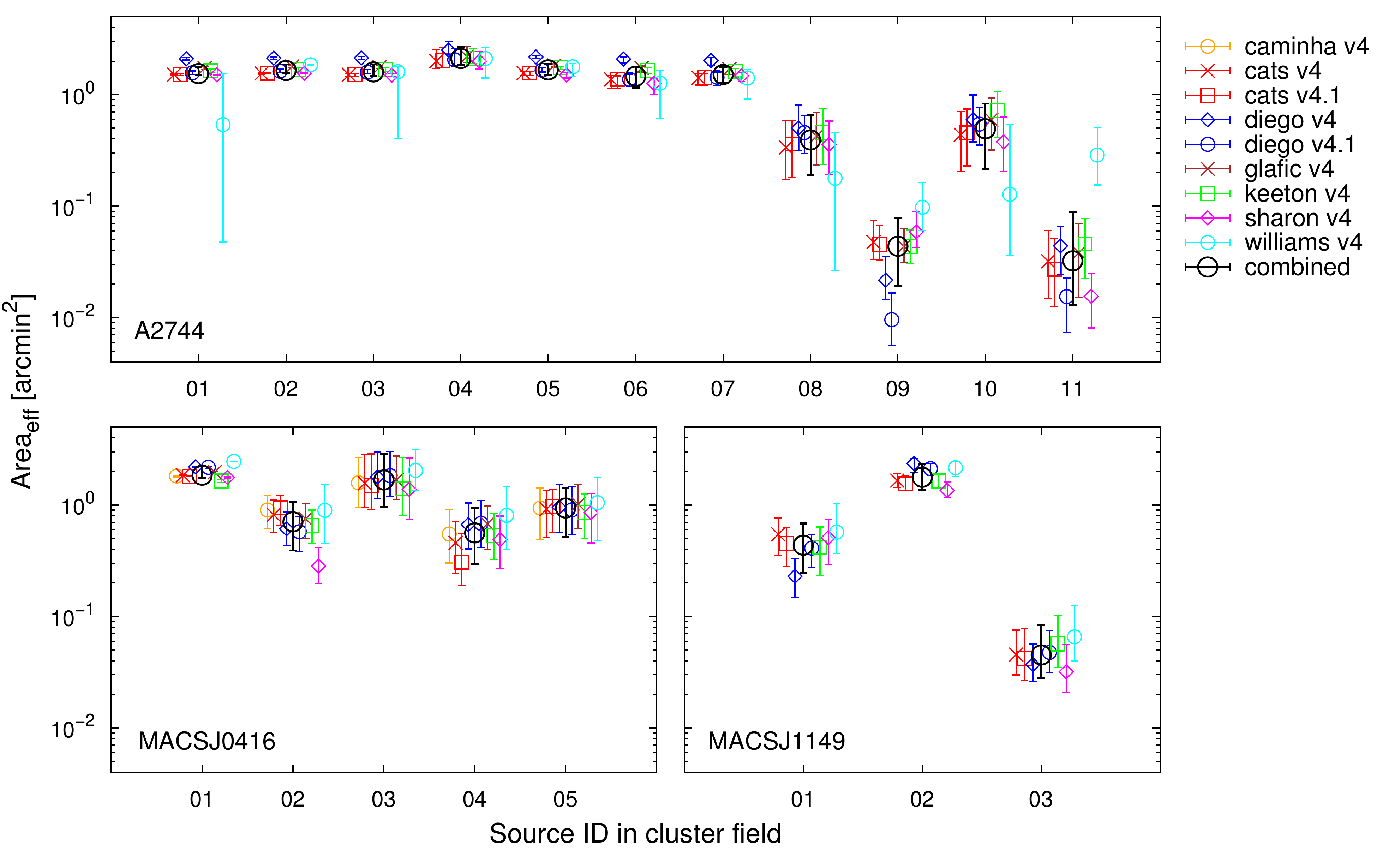}}
\caption{Median effective area per source for the lens models listed in Table \ref{tab_models} (colored symbols), and also combining all models for each cluster field (large black circles). Error bars indicate the 16th and 84th percentiles. Values for each model have been offset around the source ID for clarity.}
\label{fig_area_gal}
\end{figure*}

For computing counts, a key step is to estimate the effective area, $A_{\textmd{eff}}$, over which the source can be detected. That is, the angular area in the source plane where a map pixel having a given peak intensity can be detected over a given S/N threshold. The effective area at a given demagnified peak intensity depends not only on the PB response, but also on the source redshift assumed and the lens model adopted.

At a given redshift, we estimate the effective area as a function of demagnified peak intensity $S_{\textmd{demag,peak}}$ (corrected for PB attenuation) as follows. We consider a PB-corrected rms map for each cluster. For each ``range'' map in a given lens model, we deflect both the PB-corrected rms and magnification maps to the source plane using the deflection fields (see $\S$\ref{sect_models}). If several pixels in the image plane are deflected to only one in the source plane, only the image-plane pixel with the highest magnification is kept and assigned to the source-plane pixel (following \citealt{Coe2015}). The lensing-corrected rms level for each source-plane pixel, $\sigma_{\textmd{demag}}$, is then given by the ratio between its PB-corrected rms and magnification. At a given $S_{\textmd{demag,peak}}$, we collect all the source-plane pixels where $S_{\textmd{demag,peak}}/\sigma_{\textmd{demag}}\geq4.5$. The effective area corresponds to the sum of areas of source-plane pixels meeting this criterion, each of them given by the ALMA mosaic resolution. We precompute $A_{\textmd{eff}}$ vs $S_{\textmd{demag,peak}}$ curves for each of the redshifts used in our set of precomputed ``range'' magnification maps (see $\S$\ref{sect_models}).

For each source, we used its full distribution of demagnified peak intensities to compute its effective area. We obtain the $S_{\textmd{demag,peak}}$ distribution as in $\S$\ref{sect_fluxintr}, but using a Gaussian distribution for the image-plane peak intensity $S_{\textmd{obs,peak}}$ instead of $S_{\textmd{obs}}$. We perform a Monte Carlo simulation where we use the same number of realizations and follow the same approach for obtaining both random $S_{\textmd{obs,peak}}$ and $z$ values as in $\S$\ref{sect_magnif}. This time, however, we need to resample directly the set of ``range'' magnification maps, in such a way that the same magnification map is used for obtaining both $S_{\textmd{demag,peak}}$ and $A_{\textmd{eff}}$. This is required in order to have consistency between their values, since both depend on $\mu$ values (of the source centroid pixel and all $\textmd{PB}>0.5$ pixels, respectively) in an individual ``range'' map.

This resampling is done using the ``range'' map identifiers, which are numbered from 0 to $N_{\textmd{range}}-1$ (with $N_{\textmd{range}}$ the number of ``range'' maps provided for each model). We draw a random ``range'' map identifier using a uniform distribution bounded by zero and $N_{\textmd{range}}-1$. Using the ``range'' map corresponding to that identifier, we obtain the source magnification in the realization at the random $z$ value. We then use Eq. \ref{eq_fluxintr} for computing the source demagnified peak intensity, and then use the two closest redshift bins in our precomputed set (see $\S$\ref{sect_models}) for estimating the source effective area for that ``range'' map: first linearly interpolating precomputed $A_{\textmd{eff}}$ vs $S_{\textmd{demag,peak}}$ curves in both redshifts bins, and later linearly interpolating the $A_{\textmd{eff}}$ vs $z$ trend within these redshift limits.

The mass reconstruction for each cluster and lens model predicts a distinct proportion between high-$\mu$ and low-$\mu$ pixels at a given redshift. This is the main driver shaping the slope of the $A_{\textmd{eff}}$ vs $S_{\textmd{demag,peak}}$ curve. Finding small effective areas at low demagnified peak intensities is a natural consequence of having few regions in the maps with very high magnification. In general, the effective area increases steeply with peak intensity until some point where it reaches a plateau. In a given model, both the slope at low peak intensities and plateau level at high peak intensity depend on the modeled cluster field and adopted source redshift.

We illustrate this in Fig. \ref{fig_area_flux_z} for the CATS v4 model. At $z=2$, for instance, the largest effective areas found are $1.63_{-0.02}^{+0.02}$, $1.87_{-0.02}^{+0.01}$, and $1.79_{-0.01}^{+0.02}\,\textmd{arcmin}^2$ for A2744, MACSJ0416, and MACSJ1149, respectively. They sum to a total effective area of $\approx5.3\,\textmd{arcmin}^2$. This source-plane area is around 2.6 times smaller than the total image-plane coverage (see $\S$\ref{sect_snr5}). In the low peak intensity regime, lower source redshifts give smaller effective areas, while at $S_{\textmd{demag,peak}}\gtrsim0.2\,\textmd{mJy\,beam}^{-1}$ the opposite occurs. At $0.06-0.1\,\textmd{mJy\,beam}^{-1}$, the steepness of the $A_{\textmd{eff}}$ vs $S_{\textmd{demag,peak}}$ curves in a log-log scale are such that uncertainties of for instance $0.2\,\textmd{dex}$ in source peak intensity lead to uncertainties around $0.5\,\textmd{dex}$ in source effective area. However, the curves become shallower below $0.06\,\textmd{mJy\,beam}^{-1}$, giving a scatter in effective area of around the same order of magnitude (or below) than that in peak intensity. We find a similar qualitative behavior in the rest of the lens models used in this work, changing the numbers in the aforementioned effective areas and peak intensities.

\begin{figure}
\centering
\resizebox{0.85\hsize}{!}
{\includegraphics{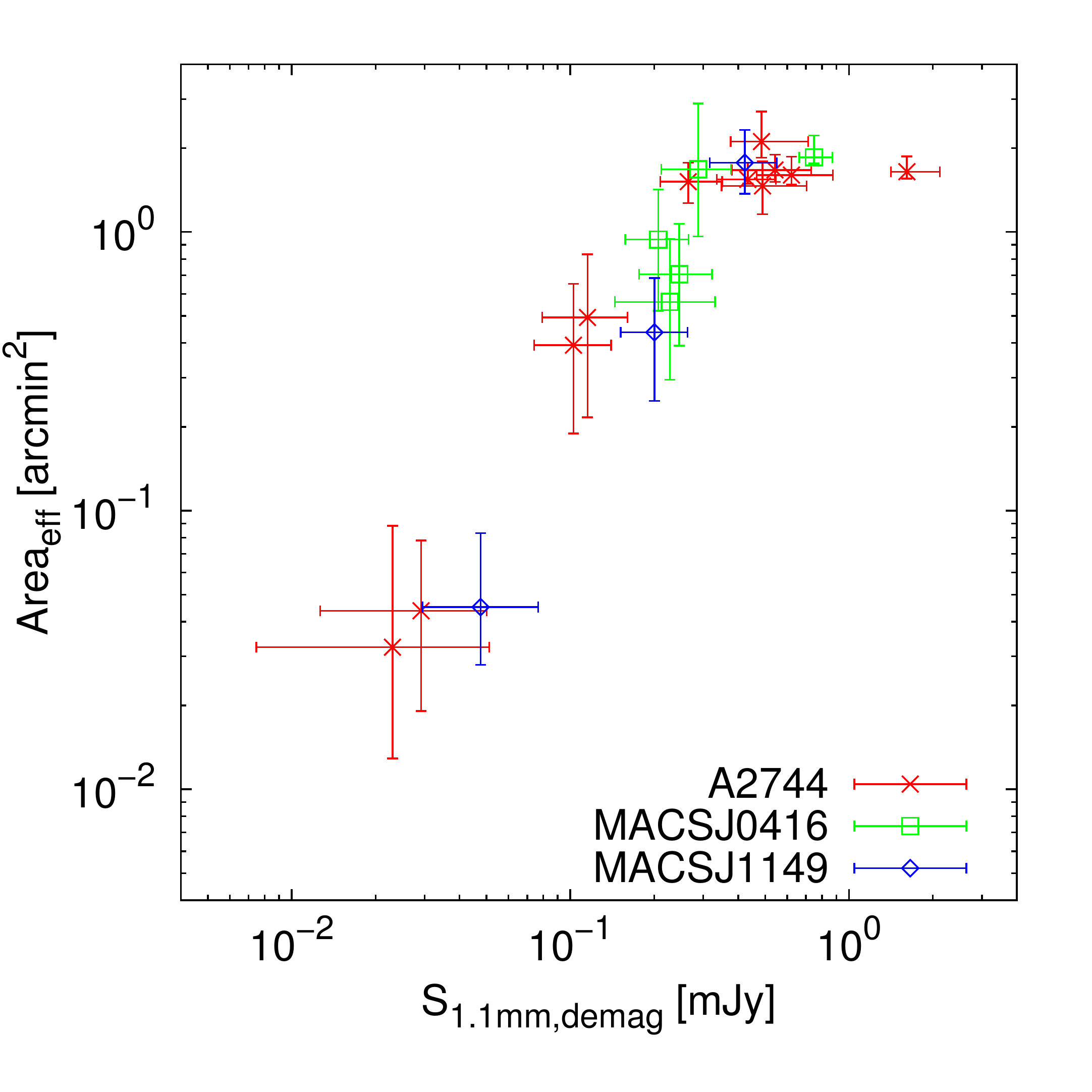}}
\caption{Median effective area as a function of demagnified integrated flux density for A2744 (red crosses), MACSJ0416 (green squares), and MACSJ1149 (blue diamonds). Median values are obtained combining all models for each cluster field. Error bars correspond to the 16th and 84th percentiles. For comparing uncertainty values, both axes cover the same interval in order of magnitude. Within the errors, both demagnified flux densities and effective areas span around 2.5 orders of magnitude.}
\label{fig_area_flux}
\end{figure}

Figure \ref{fig_area_gal} shows the median effective area for each source, computed from both the distributions obtained for each model and joining all of them for each cluster field. Median (combined) effective areas range from $\sim0.03$ to $2.1\,\textmd{arcmin}^2$. Within the uncertainties, combined effective areas cover around 2.5 orders of magnitude. In Fig. \ref{fig_area_flux}, we compare the uncertainties in the (combined) median $S_{\textmd{demag}}$ and $A_{\textmd{eff}}$ values for our sources. In the bright end ($\gtrsim0.3\,\textmd{mJy}$) we find that sources lie at the $A_{\textmd{eff}}$ plateau, thus uncertainties in effective areas are less affected by uncertainties in $S_{\textmd{demag}}$ and more by the scatter across lens models. At $\approx0.06-0.3\,\textmd{mJy}$, sources with a $S_{\textmd{demag}}$ error of such as $0.3\,\textmd{dex}$ have an $A_{\textmd{eff}}$ error close to $0.5\,\textmd{dex}$. Below $0.06\,\textmd{mJy}$, uncertainties in both of those quantities remain comparable in terms of order of magnitude, reaching even $1\,\textmd{dex}$.

\subsection{Monte Carlo simulation for source counts}\label{sect_mcsim}

\begin{figure*}
\centering
\resizebox{\hsize}{!}{\includegraphics{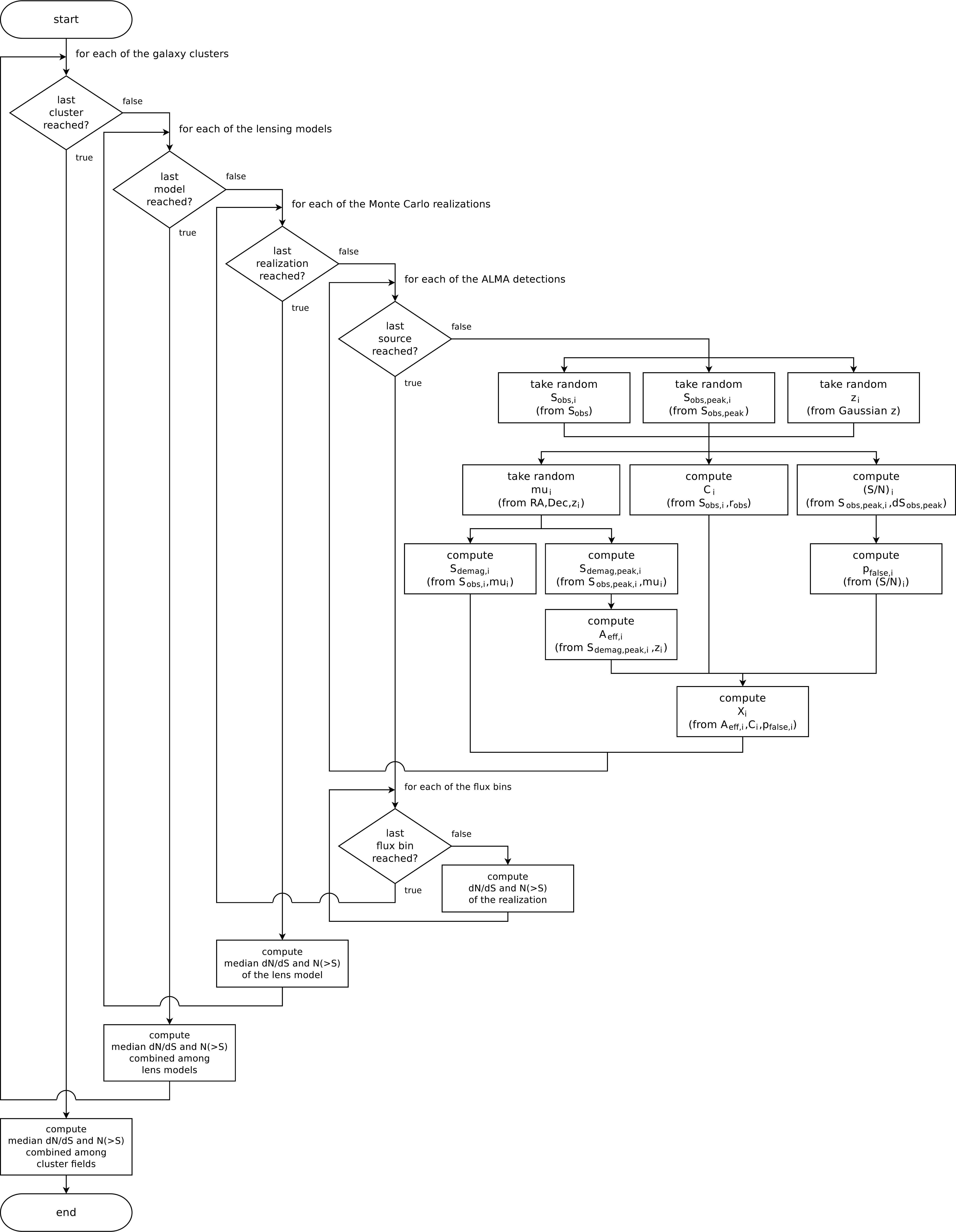}}
\caption{Diagram of the Monte Carlo simulation developed for estimating demagnified number counts (see details in $\S$\ref{sect_mcsim}).}
\label{fig_counts_flow}
\end{figure*}

We combine the techniques explained in previous sections to estimate demagnified source counts that take into account the uncertainties in observed flux densities (see Table \ref{tab_snr45gals}), adopted redshifts and modeled magnifications. We achieve this using a Monte Carlo approach. A diagram for the way in which this Monte Carlo simulation runs is shown in Fig. \ref{fig_counts_flow}. For a given galaxy cluster field and lens model, we run a total of 1000 realizations. In each of them, we compute the number counts as follows.

We generate a simulated source catalog comprised of 19 sources, keeping the same coordinates as the true detections. For each source $i$, we draw a random observed integrated flux density $S_{\textmd{obs},i}$ from a Gaussian distribution centered at $S_{\textmd{obs}}$ with standard deviation $\delta S_{\textmd{obs}}$; we proceed similarly for obtaining a random observed peak intensity $S_{\textmd{obs,peak},i}$. We also draw a random redshift from the source redshift probability distribution (see $\S$\ref{sect_z}), and use its approximated value $z_i$ as described in $\S$\ref{sect_magnif}. We then draw a random magnification $\mu_i$ as in $\S$\ref{sect_areaeff}, that is, using the identifiers of the ``range'' maps at the selected $z_i$ (and keeping a record of the selected map identifier). We also obtain the source signal-to-noise ratio as $(\textmd{S}/\textmd{N})_i=S_{\textmd{obs,peak},i}/\delta S_{\textmd{obs,peak}}$, and in the following consider only sources having $(\textmd{S}/\textmd{N})_i\geq4.5$. We then use Eq. \ref{eq_fluxintr} to obtain the demagnified integrated flux density and peak intensity ($S_{\textmd{demag},i}$ and $S_{\textmd{demag,peak},i}$) from $S_{\textmd{obs}}$, $S_{\textmd{obs,peak},i}$ and $\mu_i$. We also obtain the completeness correction $C_i$ and fraction of spurious sources $p_{\textmd{false},i}$ at the source $(\textmd{S}/\textmd{N})_i$, interpolating the curves computed in $\S$\ref{sect_comp} and $\S$\ref{sect_fsp} (see Figs. \ref{fig_comp_fluxout_reffin} and \ref{fig_fsp}). Recalling the selected map identifier at $z_i$, we obtain the effective area $A_{\textmd{eff},i}$ at the source $S_{\textmd{demag,peak},i}$ interpolating the curves precomputed in $\S$\ref{sect_areaeff}.

Having all the needed properties, we compute the differential and cumulative number counts using Eqs. \ref{eq_ndiff} to \ref{eq_xis}. We adopt $\Delta\log(S)=0.5$ and use the same flux limits for all realizations, in order to combine them later. We follow this procedure for all lens models and cluster fields. Using a given lens model, the set of realizations samples the probability distribution for the number counts in each flux bin, such that we can compute median number counts per flux bin. However, for estimating the associated uncertainties, in this case, we need to take into account low number statistics. We achieve this by computing, besides the 16th and 84th percentiles in the counts per flux bin, the Poisson confidence levels for $1\sigma$ lower and upper limits. These levels are provided by \citet{Gehrels1986} as a function of the number of events, which in our case is the median number of sources per flux bin.

We compute combined differential counts taking the median value per flux bin over the lens models listed in Table \ref{tab_models}. For computing combined cumulative counts, we take the median value per flux density limit. In both cases, we combine the counts in each cluster field separately (i.e., considering only models available for that particular field) and also combine the counts across all cluster fields.

\section{Results and discussion}\label{sect_results}

\subsection{Number counts}

\begin{figure*}
\centering
\resizebox{0.9\hsize}{!}
{\includegraphics{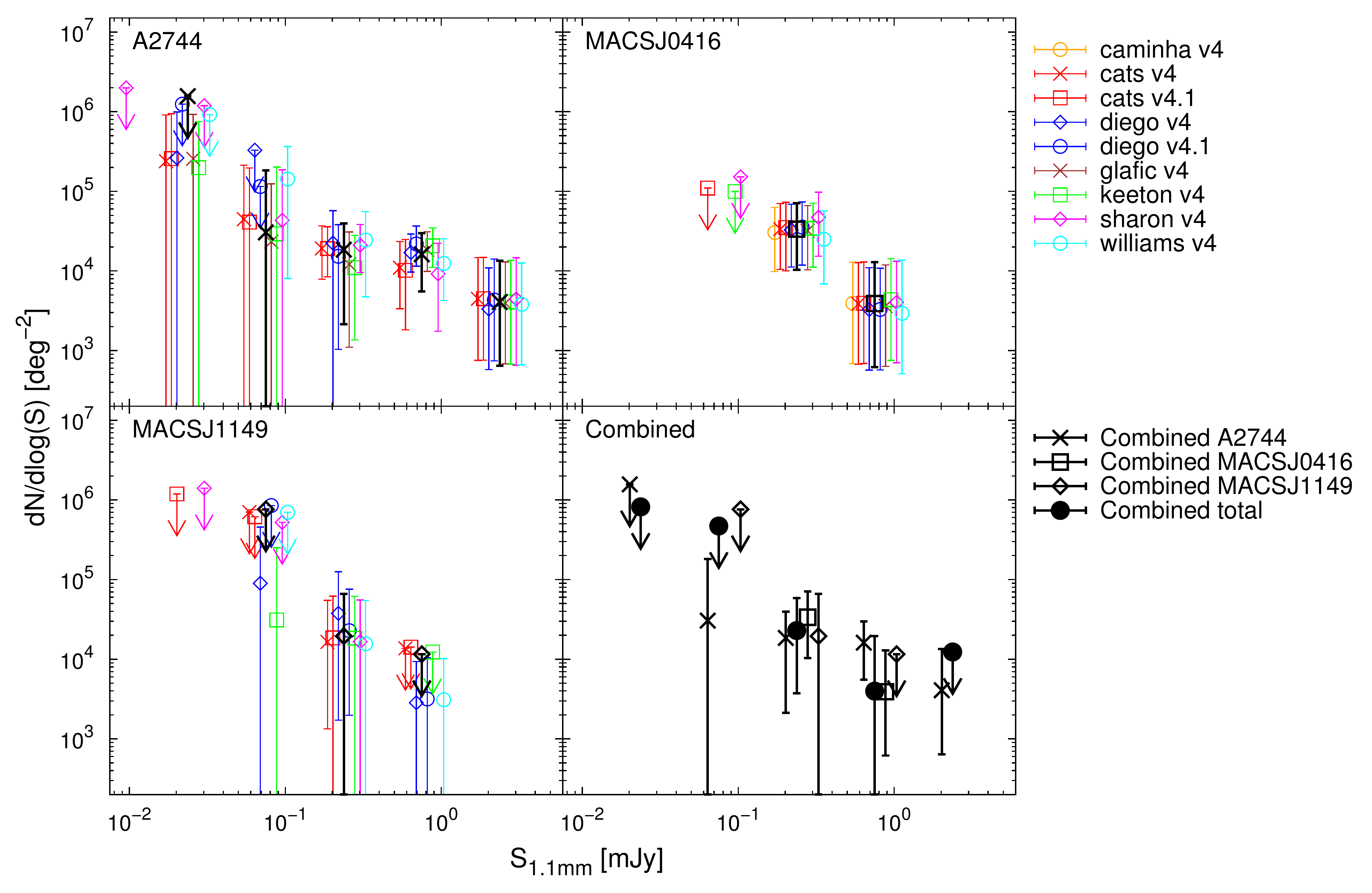}}
\caption{Demagnified differential counts at $1.1\,\textmd{mm}$, for each cluster (see legends at top-left) and combining all cluster fields (bottom-right panel). Values correspond to median counts for the lens models listed in Table \ref{tab_models} (colored symbols), combining all models for each cluster field (large black crosses, squares and diamonds) and combining all models for all cluster fields (large black filled circles). Error bars indicate the 16th and 84th percentiles, adding the scaled Poisson confidence levels for $1\sigma$ lower and upper limits respectively in quadrature. Arrows indicate $3\sigma$ upper limits for flux density bins having zero median counts and non-zero values at the 84th percentile. In the first three panels, counts for each model have been offset in flux around the combined counts for clarity. In the bottom-right panel, this is done for each galaxy cluster field around the counts that combine all models for all cluster fields.}
\label{fig_countsdiff_field_tot}
\end{figure*}

\begin{figure*}
\centering
\resizebox{0.9\hsize}{!}
{\includegraphics{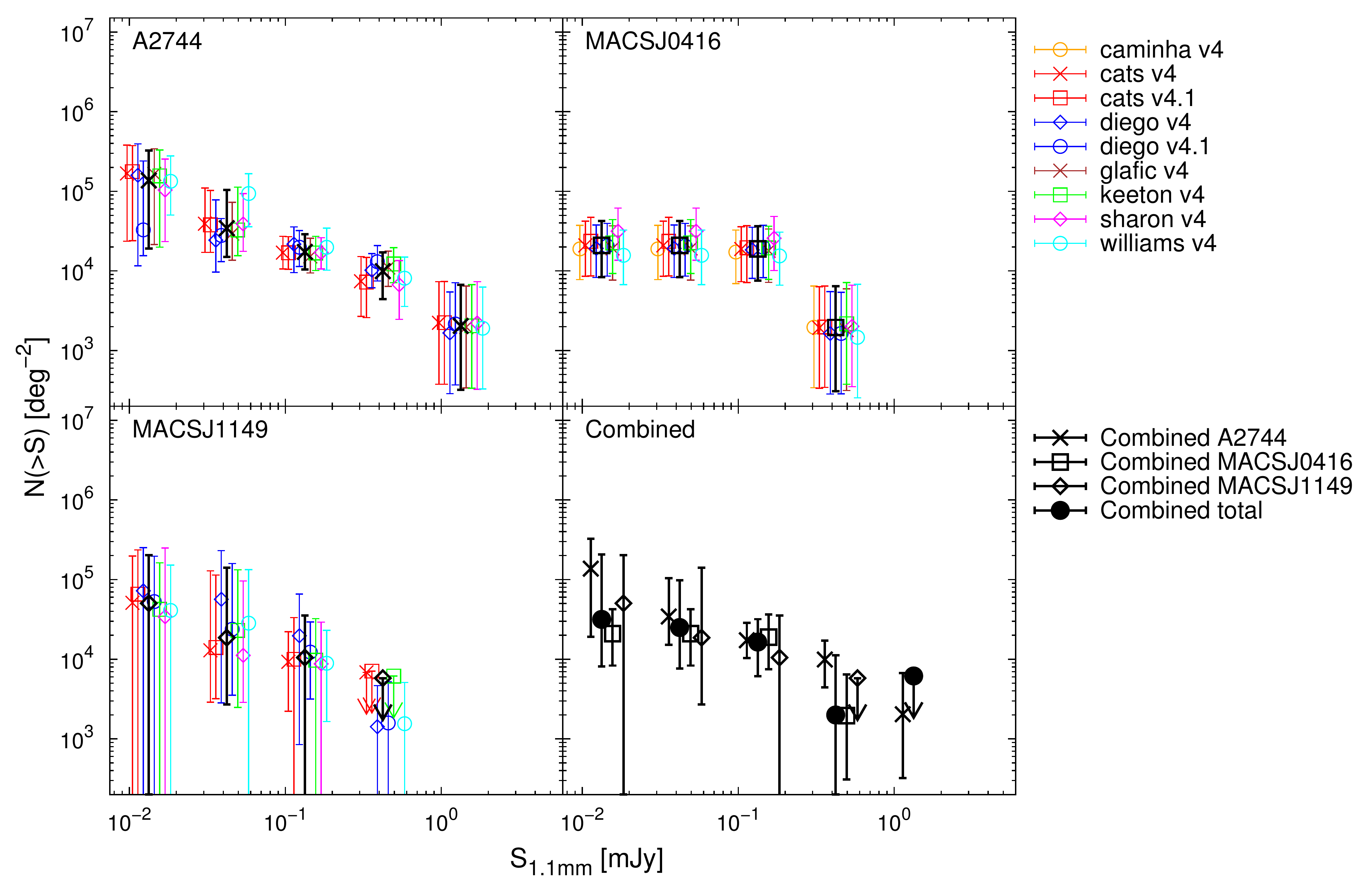}}
\caption{As in Fig. \ref{fig_countsdiff_field_tot}, but for the demagnified cumulative number counts at $1.1\,\textmd{mm}$.}
\label{fig_countscumu_field_tot}
\end{figure*}

Table \ref{tab_counts} lists the median differential and cumulative counts, combining models for each cluster field both separately and altogether. Uncertainties in the counts are obtained from the 16th and 84th percentiles, listed together with the scaled errors from $1\sigma$ lower and upper limits. When the median counts in a given flux bin are zero while having non-zero values at the 84th percentile, we only list $3\sigma$ upper limits. These counts are also presented in Figs. \ref{fig_countsdiff_field_tot} and \ref{fig_countscumu_field_tot}, where median counts for individual models in each cluster field are also displayed. Error bars shown in these Figs. combine the uncertainties from the 16th and 84th percentiles in quadrature with those from scaled Poisson confidence levels for $1\sigma$ lower and upper limits, respectively.

\begin{sidewaystable*}
\begin{center}
\caption{Demagnified $1.1\,\textmd{mm}$ number counts.}
\begin{tabular}{c|ccc|ccc}
\hline \hline
Cluster field & $S_{1.1\,\textmd{mm}}$ & $dN/d\log(S)$ & \# sources & $S_{1.1\,\textmd{mm}}$ & $N(>S)$ & \# sources \\
& $[\textmd{mJy}]$ & $[\textmd{deg}^{-2}]$ & & $[\textmd{mJy}]$ & $[\textmd{deg}^{-2}]$ & \\
\hline
A2744 & 0.024 & $<1.522\times10^6$ & $<3.0$ & 0.013 & $(1.371_{-1.091}^{+1.769}\,_{-0.446}^{+0.623})\times10^5$ & $9.0_{-1.0}^{+1.0}\,_{-2.9}^{+4.1}$ \\
& 0.075 & $(2.932_{-2.932}^{+14.30}\,_{-2.421}^{+6.710})\times10^4$ & $1.0_{-1.0}^{+1.0}\,_{-0.8}^{+2.3}$ & 0.042 & $(3.444_{-1.530}^{+6.750}\,_{-1.187}^{+1.690})\times10^4$ & $8.0_{-1.0}^{+1.0}\,_{-2.8}^{+3.9}$ \\
& 0.237 & $(1.853_{-1.241}^{+1.189}\,_{-1.005}^{+1.793})\times10^4$ & $3.0_{-2.0}^{+1.0}\,_{-1.6}^{+2.9}$ & 0.133 & $(1.733_{-0.279}^{+0.672}\,_{-0.636}^{+0.929})\times10^4$ & $7.0_{-0.0}^{+1.0}\,_{-2.6}^{+3.8}$ \\
& 0.750 & $(1.634_{-0.731}^{+0.458}\,_{-0.779}^{+1.286})\times10^4$ & $4.0_{-2.0}^{+1.0}\,_{-1.9}^{+3.1}$ & 0.422 & $(9.931_{-3.483}^{+2.485}\,_{-4.273}^{+6.685})\times10^3$ & $5.0_{-2.0}^{+1.0}\,_{-2.2}^{+3.4}$ \\
& 2.371 & $(4.090_{-0.671}^{+0.465}\,_{-3.377}^{+9.361})\times10^3$ & $1.0_{-0.0}^{+0.0}\,_{-0.8}^{+2.3}$ & 1.334 & $(2.039_{-0.341}^{+0.237}\,_{-1.684}^{+4.666})\times10^3$ & $1.0_{-0.0}^{+0.0}\,_{-0.8}^{+2.3}$ \\
\hline
MACSJ0416 & 0.024 & $0.000_{-0.000}^{+0.000}$ & $0.0_{-0.0}^{+0.0}\,_{-0.0}^{+1.8}$ & 0.013 & $(2.097_{-0.773}^{+1.364}\,_{-1.000}^{+1.650})\times10^4$ & $4.0_{-1.0}^{+1.0}\,_{-1.9}^{+3.1}$ \\
& 0.075 & $0.000_{-0.000}^{+0.000}$ & $0.0_{-0.0}^{+0.0}\,_{-0.0}^{+1.8}$ & 0.042 & $(2.097_{-0.773}^{+1.364}\,_{-1.000}^{+1.650})\times10^4$ & $4.0_{-1.0}^{+1.0}\,_{-1.9}^{+3.1}$ \\
& 0.237 & $(3.145_{-1.344}^{+1.743}\,_{-1.706}^{+3.044})\times10^4$ & $3.0_{-1.0}^{+1.0}\,_{-1.6}^{+2.9}$ & 0.133 & $(1.892_{-0.696}^{+0.916}\,_{-0.902}^{+1.489})\times10^4$ & $4.0_{-1.0}^{+1.0}\,_{-1.9}^{+3.1}$ \\
& 0.750 & $(3.915_{-0.636}^{+0.557}\,_{-3.232}^{+8.959})\times10^3$ & $1.0_{-0.0}^{+0.0}\,_{-0.8}^{+2.3}$ & 0.422 & $(1.952_{-0.318}^{+0.230}\,_{-1.611}^{+4.466})\times10^3$ & $1.0_{-0.0}^{+0.0}\,_{-0.8}^{+2.3}$ \\
& 2.371 & $0.000_{-0.000}^{+0.000}$ & $0.0_{-0.0}^{+0.0}\,_{-0.0}^{+1.8}$ & 1.334 & $0.000_{-0.000}^{+0.000}$ & $0.0_{-0.0}^{+0.0}\,_{-0.0}^{+1.8}$ \\
\hline
MACSJ1149 & 0.024 & $0.000_{-0.000}^{+0.000}$ & $0.0_{-0.0}^{+0.0}\,_{-0.0}^{+1.8}$ & 0.013 & $(5.032_{-4.121}^{+13.67}\,_{-3.241}^{+6.605})\times10^4$ & $2.0_{-1.0}^{+1.0}\,_{-1.3}^{+2.6}$ \\
& 0.075 & $<7.449\times10^5$ & $<3.0$ & 0.042 & $(1.862_{-1.046}^{+11.93}\,_{-1.199}^{+2.444})\times10^4$ & $2.0_{-1.0}^{+1.0}\,_{-1.3}^{+2.6}$ \\
& 0.237 & $(1.835_{-1.074}^{+1.306}\,_{-1.515}^{+4.199})\times10^4$ & $1.0_{-0.0}^{+0.0}\,_{-0.8}^{+2.3}$ & 0.133 & $(1.048_{-0.590}^{+0.699}\,_{-0.865}^{+2.398})\times10^4$ & $1.0_{-0.0}^{+1.0}\,_{-0.8}^{+2.3}$ \\
& 0.750 & $<1.281\times10^4$ & $<3.0$ & 0.422 & $<5.787\times10^3$ & $<3.0$ \\
& 2.371 & $0.000_{-0.000}^{+0.000}$ & $0.0_{-0.0}^{+0.0}\,_{-0.0}^{+1.8}$ & 1.334 & $0.000_{-0.000}^{+0.000}$ & $0.0_{-0.0}^{+0.0}\,_{-0.0}^{+1.8}$ \\
\hline
Combined & 0.024 & $<7.530\times10^5$ & $<3.0$ & 0.013 & $(3.157_{-1.795}^{+17.40}\,_{-1.505}^{+2.484})\times10^4$ & $4.0_{-2.0}^{+5.0}\,_{-1.9}^{+3.1}$ \\
& 0.075 & $<4.782\times10^5$ & $<3.0$ & 0.042 & $(2.501_{-1.261}^{+7.015}\,_{-1.193}^{+1.968})\times10^4$ & $4.0_{-2.0}^{+4.0}\,_{-1.9}^{+3.1}$ \\
& 0.237 & $(2.207_{-1.133}^{+1.781}\,_{-1.421}^{+2.896})\times10^4$ & $2.0_{-1.0}^{+2.0}\,_{-1.3}^{+2.6}$ & 0.133 & $(1.645_{-0.675}^{+0.828}\,_{-0.784}^{+1.294})\times10^4$ & $4.0_{-2.0}^{+3.0}\,_{-1.9}^{+3.1}$ \\
& 0.750 & $(4.025_{-1.377}^{+12.61}\,_{-3.323}^{+9.212})\times10^3$ & $1.0_{-0.0}^{+3.0}\,_{-0.8}^{+2.3}$ & 0.422 & $(1.994_{-1.994}^{+8.042}\,_{-1.646}^{+4.563})\times10^3$ & $1.0_{-1.0}^{+4.0}\,_{-0.8}^{+2.3}$ \\
& 2.371 & $<1.237\times10^4$ & $<3.0$ & 1.334 & $<6.162\times10^3$ & $<3.0$ \\
\hline
\end{tabular}
\tablefoot{Column 1: cluster field where the counts are computed. In the bottom row group, counts combining all cluster fields are listed. Column 2: flux density bin for differential counts. Column 3: median differential counts per flux bin. For non-zero median counts, uncertainties are given separately using the 16th (84th) percentiles and scaled Poisson confidence levels for $1\sigma$ lower (upper) limits. For flux density bins having zero median counts and non-zero values at the 84th percentile, only $3\sigma$ upper limits are provided. Column 4: median number of sources per flux bin. Uncertainties are given separately using the 16th (84th) percentiles and Poisson confidence levels for $1\sigma$ lower (upper) limits. Column 5: flux density limit for cumulative counts. Column 6: median cumulative counts per flux limit. Uncertainties and upper limits are as in Column 3. Column 7: median number of sources per flux limit. Uncertainties are as in Column 4.}
\label{tab_counts}
\end{center}
\end{sidewaystable*}

Because of the small number statistics, we expect our detections to give large error bars in the derived number counts. Uncertainties coming from our Monte Carlo simulation (i.e., using the whole probability distributions for observed flux densities, source redshifts and magnifications together) differ by a factor of $\sim0.05-7$ from that predicted from Poisson statistics. In A2744 and MACSJ1149, they dominate the upper error bars in the cumulative counts at flux densities below $\sim0.1\,\textmd{mJy}$ (see Table \ref{tab_counts}). This arises from the broad magnification distributions found in some of the faintest observed sources in these cluster fields. High magnifications make them the intrinsically faintest sources, with demagnified flux densities below $\sim0.1\,\textmd{mJy}$ and effective areas below $0.1\,\textmd{arcmin}^2$ (see also Fig. \ref{fig_area_flux}). For the faintest source in these cluster fields, the effective area distributions easily reach $0.03\,\textmd{arcmin}^2$ and below, which in turn elevates the counts at their flux levels. This combination of broad distributions both in demagnified flux and effective area makes the number counts below $\sim0.1\,\textmd{mJy}$ highly uncertain.

We present counts down to the flux density where at least one cluster field has non-zero combined differential counts at the 84th percentile, that is, centered on $0.007\,\textmd{mJy}$. Combining all cluster fields, our differential counts eventually span $\sim2.5$ orders of magnitude in demagnified flux density, going from the mJy level down to tens of $\mu\textmd{Jy}$. This is $\approx4$ times deeper than the observed rms level reached in our deepest ALMA FF mosaic, A2744.

We find variations across lens models in the median counts per flux bin up to $\approx1\,\textmd{dex}$. Despite this, in all cluster fields the median counts given by each model per flux bin are consistent within the error bars. A rough agreement among lens models was also found by \citet{Coe2015} when using models (at that time based on pre-FF data only) for predicting the $z>6$ NIR number counts in all the FFs. They found consistency among all models on the number of faint (i.e., at the nJy level) NIR-detected galaxies expected in HST FF observations.

We explored the effect of adopting different source redshifts in the predicted counts. Within the uncertainties, our differential counts combining all cluster fields and using redshift probability distributions according to available data (as above) are consistent with those obtained assuming a Gaussian redshift distribution centered at $z=2\pm0.5$ for all detections. We also obtain consistent results adopting exactly $z=2$ for all detections, as well as when assuming a uniform redshift distribution between the cluster redshift and $z=4$. In these three cases, variations in the median counts combining all cluster fields are only up to $\approx0.04\,\textmd{dex}$ below $1.3\,\textmd{mJy}$. Our combined counts are also in agreement within the errors with those obtained centring the Gaussian at $z=3\pm0.5$ for all detections (although upper error bars assuming this higher redshift center are greater by $\approx0.25\,\textmd{dex}$ at $\lesssim0.1\,\textmd{mJy}$, due to the larger high-magnification regions for this redshift).

\subsection{Comparison to previous ALMA studies and galaxy formation model predictions}\label{sect_counts_comp}

\begin{figure*}
\centering
\resizebox{0.7\hsize}{!}
{\includegraphics{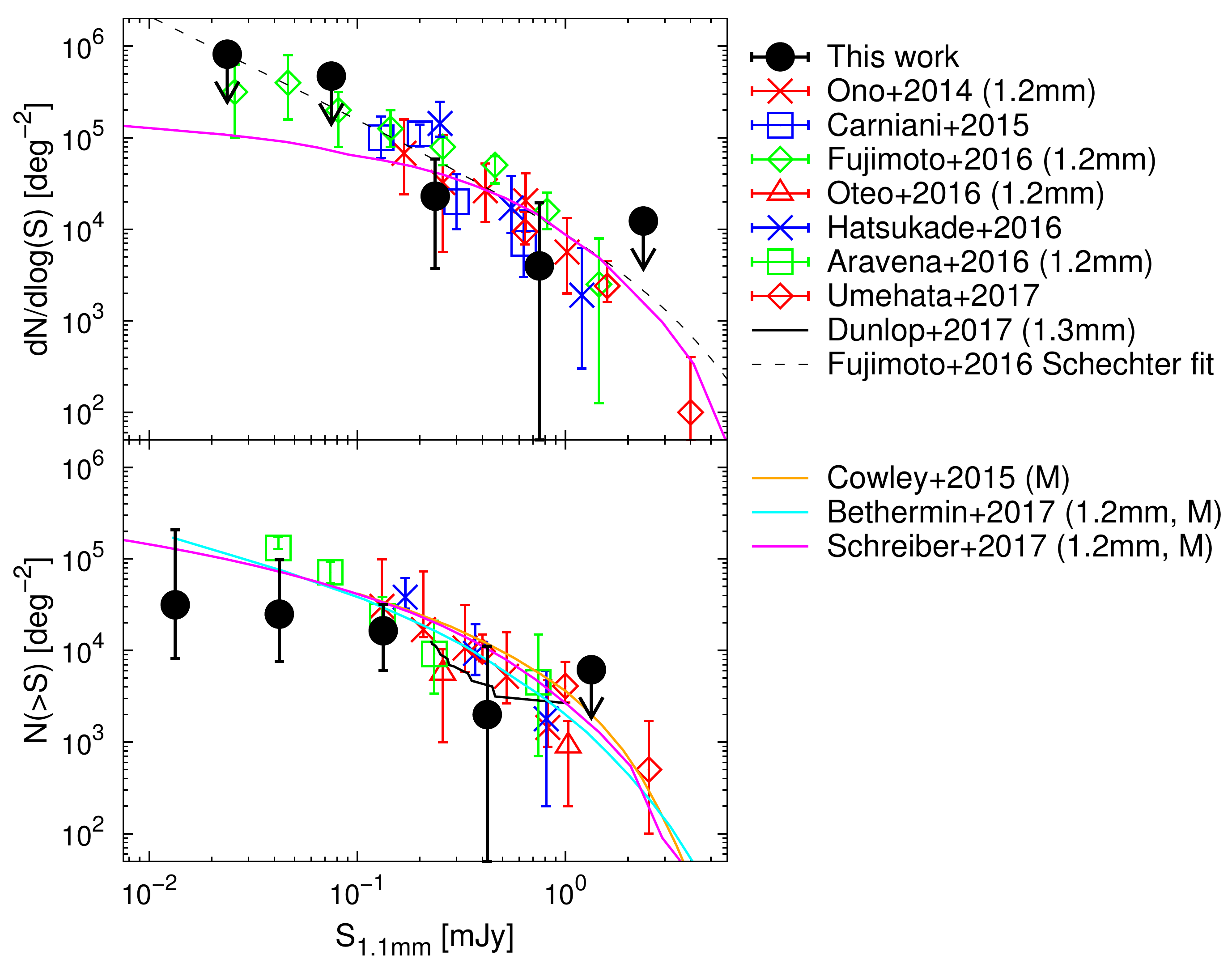}}
\caption{Differential (top) and cumulative (bottom) counts at $1.1\,\textmd{mm}$ compared to ALMA results and galaxy formation model predictions from the literature. Our counts (large black filled circles) correspond to median values combining all models for all cluster fields. Error bars indicate the 16th and 84th percentiles, adding the scaled Poisson confidence levels for $1\sigma$ lower and upper limits respectively in quadrature. Arrows indicate $3\sigma$ upper limits for flux densities having zero median counts and non-zero values at the 84th percentile. We show previous results reported by \citet{Ono2014} as red crosses, \citet{Carniani2015} as blue squares, \citet{Fujimoto2016} as green diamonds (with their Schechter fit shown as a black dashed line), \citet{Oteo2016} as red triangles, \citet{Hatsukade2016} as blue crosses, \citet{Aravena2016} as green squares, \citet{Umehata2017} as red diamonds and \citet{Dunlop2017} as a black solid curve. We show number counts predicted by the galaxy formation models from \citet{Cowley2015} (orange line), \citet{Bethermin2017} (cyan line) and \citet{Schreiber2017} (magenta line). We scale the counts derived at other wavelengths as $S_{1.1\,\textmd{mm}}=1.29\times S_{1.2\,\textmd{mm}}$ and $S_{1.1\,\textmd{mm}}=1.48\times S_{1.3\,\textmd{mm}}$ (following \citealt{Hatsukade2016}).}
\label{fig_countsdiffcumu_comp}
\end{figure*}

Figure \ref{fig_countsdiffcumu_comp} shows our $1.1\,\textmd{mm}$ number counts compared to results from recent ALMA observations that probe down to the sub-mJy level. These include counts derived from sources detected by serendipitous \citep{Ono2014,Carniani2015,Fujimoto2016,Oteo2016} as well as dedicated surveys in blank fields \citep{Hatsukade2016,Aravena2016,Dunlop2017} and around a $z=3.09$ protocluster \citep{Umehata2017}. It should be noted that these previous works use their own source detection criterion, as well as their own choice and methodology for computing corrections to the counts (e.g., completeness, flux deboosting, fraction of spurious sources, effective areas, magnifications). Recalculating their counts matching our criteria, which would ensure a fair comparison, is beyond the scope of this work. Instead, we only apply a scaling for previous counts derived at wavelengths other than $1.1\,\textmd{mm}$. In those cases, we scale their estimates as $S_{1.1\,\textmd{mm}}=1.29\times S_{1.2\,\textmd{mm}}$ and $S_{1.1\,\textmd{mm}}=1.48\times S_{1.3\,\textmd{mm}}$. These conversion factors are derived by assuming a characteristic modified blackbody spectrum (following \citealt{Hatsukade2016}), and are adopted for consistency with previous works (which assume distinct SED templates).

Within the uncertainties, our estimates for both differential and cumulative number counts are in good agreement with all the aforementioned works for the two or three brightest bins, that is, down to $0.422\,\textmd{mJy}$. At $0.133-0.422\,\textmd{mJy}$, our number counts are consistent within $1\sigma$ with all but \citet{Fujimoto2016} and \citet{Hatsukade2016} data. At flux densities fainter than $0.133\,\textmd{mJy}$, however, the derived $3\sigma$ upper limits to our differential counts are consistent with both the \citet{Fujimoto2016} data and their \citet{Schechter1976} best-fitting function. Also below this flux density, our cumulative counts are lower by $\approx1\,\textmd{dex}$ than \citet{Aravena2016} data, being inconsistent with their results at a $1\sigma$ level. Our findings suggest a flattening of the number counts.

The counts derived from serendipitously detected sources are based on detections in fields that targeted a previously defined set of sources. These counts are expected to be biased, as the detections might be clustered around the original targets \citep{Hatsukade2016}. Restricting only to flux densities above $0.133\,\textmd{mJy}$, we are not able to quantify this bias, given the large uncertainties in our derived counts. Neither can we make a strong distinction between our counts, which are based solely on observations lensed  by galaxy clusters, and those derived from blank-field observations. Intriguingly, our counts in the brighter flux density bins are consistent with those found by \citet{Umehata2017} toward the SSA22 protocluster, both including and not including their detections having spectroscopic redshifts coincident with the protocluster (in Fig. \ref{fig_countsdiffcumu_comp} we show only the first case).

In addition to recent ALMA data, Fig. \ref{fig_countsdiffcumu_comp} shows the counts predicted by galaxy formation models down to $\lesssim0.1\,\textmd{mJy}$. In particular, \citet{Cowley2015} use the semi-analytic model GALFORM \citep{Lacey2016} to predict the submm counts. We compare our results to their cumulative number counts at $1.1\,\textmd{mm}$ for their simulated lightcones down to $0.1\,\textmd{mJy}$. On the other hand, \citet{Bethermin2017} and \citet{Schreiber2017} present simulations of the extragalactic sky (SIDES and EGG, respectively). For \citet{Bethermin2017}, we compare our results to their cumulative number counts at $1.2\,\textmd{mm}$ for their ``intrinsic'' simulation down to $0.01\,\textmd{mJy}$, while for \citet{Schreiber2017} we compare to their differential and cumulative number counts at $1.2\,\textmd{mm}$ down to $10^{-8}\,\textmd{mJy}$. In these last two cases we rescale their counts to $1.1\,\textmd{mm}$ as was done for $1.2\,\textmd{mm}$ observations above.

\citet{Cowley2015} obtain the dust SED per galaxy in a self-consistent way (see also \citealt{Lacey2016}), using a simplified model based in the spectrophotometric code GRASIL \citep{Silva1998} that agrees with GRASIL predictions at rest-frame wavelengths $>70\,\mu\textmd{m}$. We note that the constraints for their model parameters include the observed cumulative counts at $850\,\mu\textmd{m}$ and the redshift distribution of sources with flux density $>5\,\textmd{mJy}$ at $850\,\mu\textmd{m}$ (see \citealt{Lacey2016}). \citet{Bethermin2017} and \citet{Schreiber2017} use the phenomenological model 2SFM (two star formation modes; \citealt{Sargent2012}), which is based on the observed evolution of the main sequence with redshift. Both groups add their own assumptions for constructing the mock catalogs and estimating further source properties from empirical prescriptions. They also choose particular SED libraries (which cover the FIR-to-submm wavelengths) for assigning spectra to model sources based on these properties (see \citealt{Bethermin2017,Schreiber2017}). Both groups calibrate their models using particular sets of observational constraints for the SED evolution, based on stacking analyses. For \citet{Bethermin2017}, these constraints include LABOCA $870\,\mu\textmd{m}$ and AzTEC $1.1\,\textmd{mm}$ data (see \citealt{Bethermin2015b}). \citet{Schreiber2017} note that at $1.2\,\textmd{mm}$ they do not calibrate their FIR SEDs nor prescriptions, although their constraints include ALMA $870\,\mu\textmd{m}$ data in the Extended Chandra Deep Field South (Extended CDFS). From Fig. \ref{fig_countsdiffcumu_comp}, we note that these three models agree well in the displayed flux range. Although none of the models predict number counts as shallow as our estimates, at flux densities $<0.1\,\textmd{mJy}$ they predict counts around $1\sigma$ lower than \citet{Fujimoto2016} and \citet{Aravena2016} values, and agree with our predictions within $1\sigma$ uncertainties.

\subsection{Effect of source sizes}

\begin{figure}
\centering
\resizebox{0.85\hsize}{!}{\includegraphics{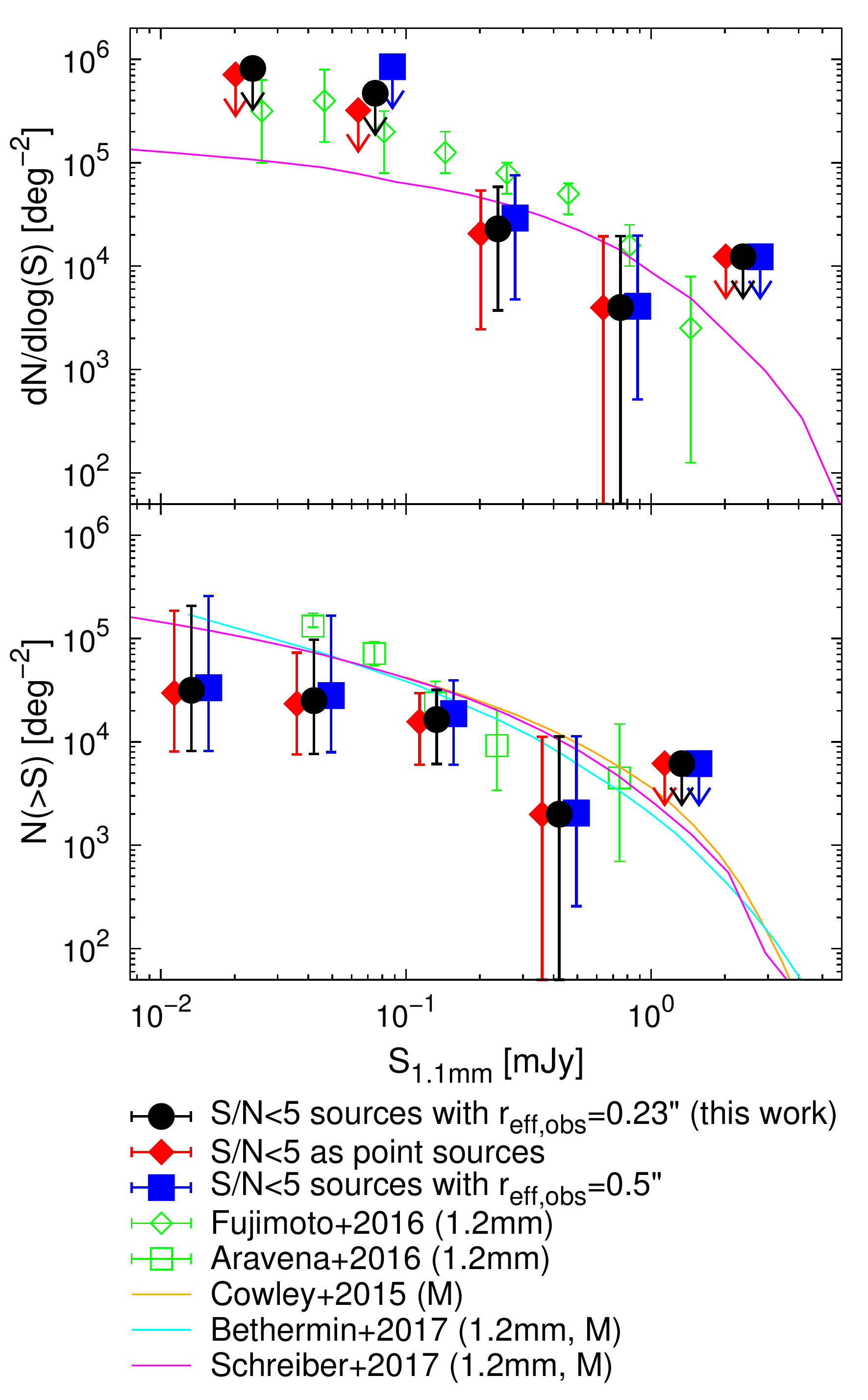}}
\caption{Differential (top) and cumulative (bottom) counts at $1.1\,\textmd{mm}$ for different assumptions regarding the image-plane source scale radii for low-significance sources: adopting $r_{\textmd{eff,obs}}=0\farcs23$ (black filled circles, fiducial); assuming they are point sources (red filled diamonds); and adopting $r_{\textmd{eff,obs}}=0\farcs5$ (blue filled squares). Our counts correspond to median values combining all models for all cluster fields. Error bars indicate the 16th and 84th percentiles, adding the scaled Poisson confidence levels for $1\sigma$ lower and upper limits respectively in quadrature. Arrows indicate $3\sigma$ upper limits for flux densities having zero median counts and non-zero values at the 84th percentile. We show previous results reported by \citet{Fujimoto2016} as green diamonds and \citet{Aravena2016} as green squares. We show number counts predicted by the galaxy formation models from \citet{Cowley2015} (orange line), \citet{Bethermin2017} (cyan line) and \citet{Schreiber2017} (magenta line). We scale the counts derived at other wavelength as $S_{1.1\,\textmd{mm}}=1.29\times S_{1.2\,\textmd{mm}}$ (following \citealt{Hatsukade2016}).}
\label{fig_countsdiffcumu_reff}
\end{figure}

For exploring the effect of varying the image-plane source sizes on the demagnified number counts, we test the following extreme cases, which should bracket our expectations. Firstly, assuming that $4.5\leq\textmd{S/N}<5$ sources are point sources. This assumption is supported by recent results regarding DSFG sizes, both from several publicly available ALMA maps at $1\,\textmd{mm}$ \citep{Fujimoto2017} and an ALMA follow-up of SCUBA2 sources in the CDFS at $850\,\mu\textmd{m}$ (Gonz\'alez-L\'opez et al. 2018, in prep.). \citet{Fujimoto2017} find a positive correlation between source size and bolometric luminosity in the FIR; an extrapolation of this trend to lower luminosities may suggest that our sources at lower S/N are more compact than high-significance detections. Similarly, Gonz\'alez-L\'opez et al. (2018, in prep.) find that robustly selected DSFGs at a few mJy (equivalent to $\gtrsim0.5\,\textmd{mJy}$ at $1.1\,\textmd{mm}$) in the CDFS have compact sizes on average, with a median effective radius $\approx0\farcs08$. And secondly, assuming that $4.5\leq\textmd{S/N}<5$ sources have more extreme observed effective radii of $r_{\textmd{eff,obs}}=0.5''$. This value is motivated by the $1\sigma$ uncertainty found for the largest image-plane scale radius among the high-significance detections. In this case, we obtain the integrated flux densities of the low-significance detections scaling the peak intensities by the typical ratios 0.55, 0.84, and 0.84 in A2744, MACSJ0416, and MACSJ1149, respectively.

Estimated number counts for these cases are shown in Fig. \ref{fig_countsdiffcumu_reff}, together with our fiducial case. We find that assuming $r_{\textmd{eff,obs}}=0.5''$ for low-significance sources leads to an agreement with \citet{Aravena2016} at $1\sigma$. Assuming that our low-significance detections are point sources disagrees with their estimates at $3\sigma$, although remains consistent with \citet{Fujimoto2016} counts assuming our $3\sigma$ upper limit.

Below $0.133\,\textmd{mJy}$, our fiducial number counts are consistent with available data from both serendipitous and blank-field surveys only at a $3\sigma$ level. The discrepancy with our median counts could be attributed to cosmic variance or to the aforementioned observational biases. However, it may also reveal the need for further corrections in our number counts. More specifically, we may require a proper treatment for the stretching that source shapes experience in the image plane. There may be sources missed because their high magnifications led them to have lensed angular sizes greater than a synthesized beam. In order to take these effects into account, we would need to assume a distribution of source sizes at several redshifts and a set of different intrinsic source geometries, as well as passing them through the $uv$ and lens modeling.  Accounting for this could elevate the derived number counts in the faint end, if the dust emission from low-significance detections is more extended than suggested by extrapolations to current observational data (\citealt{Fujimoto2017} and Gonz\'alez-L\'opez et al. 2018 in prep., see $\S$\ref{sect_snr45}). We leave a detailed analysis about the impact of demagnified source geometries on the number counts  for future work, so at low flux densities our reported counts are strictly computed for the beam size quoted for each ALMA mosaic.

\subsection{Contribution to the extragalactic background light}

We use the Monte Carlo realizations of the differential number counts to compute the contribution to the EBL provided by each of them, adding up the contribution contained in each flux bin. From this procedure, we estimate a median contribution of $8.222_{-4.188}^{+8.837}$ ($3.861_{-2.059}^{+7.847}$) $\textmd{Jy}\,\textmd{deg}^{-2}$ resolved in our demagnified sources at $1.1\,\textmd{mm}$ down to $0.013$ ($0.133$) $\textmd{mJy}$, with uncertainties computed from the 16th and 84th percentiles.

We compare our estimate with the total EBL measurement at that wavelength estimated by the Planck collaboration using their best-fit extended halo model. Following \citet{Aravena2016}, we interpolate the Planck estimate (see \citealt{Ade2014}, Table 10) finding an EBL of $19.143_{-0.723}^{+0.751}\,\textmd{Jy}\,\textmd{deg}^{-2}$ at $263.14\,\textmd{GHz}$, which is the set Local Oscillator frequency for our observations (see Paper I). The contribution provided by our demagnified sources represents $43_{-22}^{+46}\%$ ($20_{-11}^{+41}\%$) of this EBL at $1.1\,\textmd{mm}$ down to $0.013$ ($0.133$) $\textmd{mJy}$. As expected from Fig. \ref{fig_countsdiffcumu_comp}, this contribution is lower than (although consistent to $\approx1.5\sigma$ with) results by \citet{Carniani2015} and \citet{Hatsukade2016}, both at $1.1\,\textmd{mm}$. \citet{Carniani2015} found an estimate of $17_{-5}^{+10}\,\textmd{Jy}\,\textmd{deg}^{-2}$ down to $0.1\,\textmd{mJy}$, while a value around 12 (14) $\textmd{Jy}\,\textmd{deg}^{-2}$ is obtained when we extrapolate the Schechter (double power law) best-fitting function by \citet{Hatsukade2016} down to $0.1\,\textmd{mJy}$.

\section{Summary}\label{sect_concl}

We have derived lensing-corrected number counts at $1.1\,\textmd{mm}$ exploiting: 1) the high resolution and depth reached in a dedicated ALMA survey of three galaxy clusters (i.e., A2744, MACSJ0416, and MACSJ1149) as part of the Frontier Fields program, and 2) the public availability of several models for the mass reconstruction of these clusters. This is the first time that the surface density of DSFGs is estimated around three well-studied galaxy clusters using ALMA data. Our source catalog includes $\textmd{S/N}\geq5$ detections already introduced with the ALMA Frontier Fields Survey (Paper I), plus $4.5\leq\textmd{S/N}<5$ detections reported in the present work. We correct the counts for completeness and fraction of spurious sources. Moreover, we develop a careful treatment to fold the magnification uncertainties in the derived counts using a Monte Carlo simulation.

Our ALMA mosaics of the three FF galaxy clusters cover a total observed area of $\sim14\,\textmd{arcmin}^2$, which results in a smaller effective area in the source plane once a lens model is applied (e.g., the total area is reduced by $\sim2.6$ times in the CATS v4 model for a source-plane $z=2$). Combining all cluster fields, our differential number counts span $\sim2.5$ orders of magnitude in demagnified flux density, going from the mJy level down to tens of $\mu\textmd{Jy}$. We find an overall agreement between the counts derived for different lens models in a given cluster field. Within the error bars in our number counts (coming from both Poisson errors and lensing model uncertainties) our results are consistent at $3\sigma$ with recent estimates from deep ALMA observations \citep{Ono2014,Carniani2015,Fujimoto2016,Oteo2016,Hatsukade2016,Aravena2016,Umehata2017,Dunlop2017}. However, below $\approx0.1\,\textmd{mJy}$ our cumulative number counts are $\approx1\,\textmd{dex}$ lower than previous estimates. Our work suggests a flattening of the number counts, and implies that we may finally be seeing a turn over.

Using publicly available lens models and a statistical approach, we are able to derive $1.1\,\textmd{mm}$ number counts around three galaxy clusters, down to demagnified flux densities $\approx4$ times fainter than the rms level reached in our deepest ALMA mosaic. This highlights the potential of finding even fainter sources in these FFs with deeper ALMA data, suggesting that future $1.1\,\textmd{mm}$ observations reaching an rms of such as $10\,\mu\textmd{Jy}$ could yield number counts down to $\approx2.5\,\mu\textmd{Jy}$ in these fields. Additionally, further spectroscopic redshift determinations for our detections could serve as new constraints for lensing models, helping to increase the accuracy in the magnification estimates \citep{Johnson2016} and hence in the number counts derived from future deep surveys.

\begin{acknowledgements}

We thank the referee for the comments and suggestions which contributed to improve this paper. We acknowledge support from CONICYT through FONDECYT grant 3160776 (A.M.M.A.), CONICYT grants BASAL-CATA PFB-06/2007 (F.E.B), FONDECYT Regular 1141218 (F.E.B., J.G.-L.), Programa de Cooperaci\'on Cient\'ifica ECOS-CONICYT C16U02 (F.E.B., J.G.-L.), Programa de Astronom\'ia FONDO ALMA 2016 31160033 (J.G.-L.), CONICYT + Programa de Astronom\'ia + Fondo CHINA-CONICYT (J.G.-L.), the Ministry of Economy, Development, and Tourism's Millennium Science Initiative through grant IC120009, awarded to The Millennium Institute of Astrophysics, MAS, Chile (C.R.-C. and F.E.B.) and CONICYT through FONDECYT grant 3150238 (C.R.-C.). E.I. acknowledges partial support from FONDECYT through grant 1171710. R.D. gratefully acknowledges the support provided by the BASAL Center for Astrophysics and Associated Technologies (CATA). This work is partly sponsored by the Chinese Academy of Sciences (CAS), through a grant to the CAS South America Center for Astronomy (CASSACA) in Santiago, Chile (C.R.-C.). The ALMA observations were carried out under program ADS/JAO.ALMA\#2013.1.00999.S. ALMA is a partnership of ESO (representing its member states), NSF (USA) and NINS (Japan), together with NRC (Canada) and NSC and ASIAA (Taiwan), in cooperation with the Republic of Chile. The Joint ALMA Observatory is operated by ESO, AUI/NRAO and NAOJ. This work utilizes gravitational lensing models produced by PIs Brada{\v c}, Natarajan \& Kneib (CATS), Merten \& Zitrin, Sharon, Williams, Keeton, Bernstein and Diego, and the GLAFIC group. This lens modeling was partially funded by the HST Frontier Fields program conducted by STScI. STScI is operated by the Association of Universities for Research in Astronomy, Inc. under NASA contract NAS 5-26555. The lens models were obtained from the Mikulski Archive for Space Telescopes (MAST).

\end{acknowledgements}

\end{document}